\documentclass[12pt,notitlepage,fleqn]{article}
\usepackage{fancyhdr}
\usepackage{amsfonts}
\usepackage{graphicx}
\usepackage{amsmath}
\usepackage{indentfirst}

\input{tcilatex}
\begin{document}

\begin{center}
{\Huge \ \ Using Phenomenological Formulae,}

{\Huge Deducing the Masses and Flavors of Quarks, Baryons and Mesons from
Two Elementary Quarks }

\ \ \ \ \ \ \ \ \ \ \ \ \ \ \ \ \ \ \ \ \ \ \ \ \ \ \ \ \ \ \ \ \ \ \ \ \ \
\ \ \ \ \ \ \ \ \ \ \ \ \ \ \ \ \ \ \ \ \ \ \ \ \ \ \ \ 

{\normalsize Jiao-Lin Xu}

\ \ \ \ \ \ \ \ \ \ \ \ \ \ \ \ \ \ \ \ \ \ \ \ \ \ \ \ \ \ \ \ \ \ \ 

{\small The Center for Simulational Physics, The Department of Physics and
Astronomy}

{\small University of Georgia, Athens, GA 30602, USA}

e-mail: {\small \ jxu@hal.physast.uga.edu}

\bigskip

\ \ \ \ \ \ \ \ \ \ \ \ \ \ \ \ \ \ \ \ \ \ \ \ \ \ \ \ \ \ \ \ \ \ \ \ \ \
\ \ \ \ \ \ \ \ \ \ \ \ \ \ \ \ \ \ \ \ \ \ \ \ \ \ \ \ \ \ \ \ \ \ \ \ \ \
\ \ \ \ \ \ \ \ \ \ \ \ \ \ \ \ \ \ \ \ \ \ \ \ \ \ \ \ \ \ \ \ \ \ \ \ \ \
\ \ \ 

\ \ \ \ \ \ \ \ \ \ \ \ \ \ \ \ \ \ \ \ \ \ \ \ \ \ \ \ \ \ \ \ \ \ \ \ \ \
\ \ \ \ \ \ \ \ \ \ \ \ \ \ \ \ \ \ \ \ \ \ \ \ \ \ \ \ \ \ \ \ \ \ \ \ \ \
\ \ \ \ \ \ \ \ \ \ 

\textbf{Abstract}

\ \ \ \ \ \ \ \ \ \ \ \ \ \ \ \ \ \ \ \ \ \ \ \ \ \ \ \ \ \ \ \ \ \ \ \ \ \
\ \ \ \ \ \ \ \ \ \ \ \ \ \ \ \ \ \ \ \ \ \ \ \ \ \ \ \ \ \ \ \ \ \ \ \ \ \
\ \ \ \ \ \ \ \ \ \ \ \ \ \ \ \ \ \ 
\end{center}

{\small \ \ Using phenomenological formulae, we deduce the masses and
quantum numbers of the quarks from two elementary quarks (}$\epsilon _{u}$ 
{\small and }$\epsilon _{d}$) {\small first. Then using the sum laws and a\
binding energy formula, in terms of the qqq baryon model and SU(4), we
deduce the masses and quantum numbers of the important baryons from the
deduced quarks. At the same time, using the sum laws and a\ binding energy
formula, in terms of the quark-antiquark bound state meson model, we deduce
the masses and quantum numbers of the mesons from the deduced quarks. The
deduced masses of the baryons and mesons are 98\% consistent with
experimental results. The deduced quantum numbers of the baryons and mesons
match with the experimental results exactly. This paper improves upon the
Quark Model, making it more powerful and more reasonable. It predicts some
baryons also.\ PACS: 12.39.-x; 14.65.-q; 14.20.-c\ \ \ keywords:
phenomenology, elementary quark, mass, SU(4), baryon, meson\ \ \ }

\ \ \ \ \ \ \ \ \ \ \ \ \ \ \ \ \ \ \ \ \ \ \ \ \ \ \ \ \ \ \ \ \ \ \ \ \ \
\ \ 

\ \ \ \ \ \ \ \ \ \ \ \ \ \ \ \ \ \ \ \ \ \ \ \ \ \ \ \ \ \ \ \ \ \ \ \ \ \
\ \ \ \ \ \ \ \ \ \linebreak \textbf{1. \ Introduction}{\small \ \ \ \ \ \ }

\ \ \ \ \ \ \ \ \ \ \ \ \ \ \ \ \ \ \ \ \ \ \ \ \ \ \ \ \ \ \ \ \ \ \ \ \ \
\ \ \ \ \ \ 

Using SU(3)$_{\text{f}}$, SU(4)$_{\text{f}}$ and SU(5)$_{\text{f}}$, the
Quark Model \cite{Quark Model} has successfully deduced the quantum numbers
of baryons and mesons. It also has found some masses of new baryons or
mesons, from the masses of the old discovered baryons or mesons, using mass
relations of Gell-Mann--Okubo mass formulae \cite{Gell-Mann-Okubo}. Thus the
Quark Model has led to many discoveries of baryons and mesons.\ These works
are some of the greatest works in particle physics. There are, however, the
important problems that need to solve:

(1) The quark's intrinsic quantum numbers cannot deduce and have to put in
\textquotedblleft by hand\textquotedblright\ \cite{By hand}. How do we
deduce the flavor of the flavored quarks? Why do all quarks have the same
spin (s = $\frac{1}{2}$) and baryon number (B = $\frac{1}{3})$ ?

(2) Are the quarks really all elementary particles? Why the quark with large
mass always automatically decays into a smaller mass quark in a very short
time?

(3) Why the different flavored quarks have the flavor symmetry SU(3), SU(4)
and SU(5)? What is the physical foundation of the flavor symmetry?

(4) How do we deduce the mass spectrum of the quarks? the quark masses, as
arbitrary parameters, led to that the standard model has too many arbitrary
parameters (nineteen) and thus the standard model is incomplete\ \cite%
{Incomplete}.

(5) How do we deduce the mass spectrums of the baryons and the mesons?

These problems are concerned with the physical foundation of the standard
model. We can not dodge these problems. There may be some clues about new
theory in the solutions of these problems. Using phenomenological formulae,
this paper deduces the masses and quantum numbers (including the flavor
numbers S, C and B) of the current quarks from the two kinds of elementary
quarks $\epsilon _{u}$ and $\epsilon _{d}$. This paper shows that all
current quarks inside hadrons are the excited states of $\epsilon _{u}$ or $%
\epsilon _{d}$. Since the two elementary quarks ($\epsilon _{u}$ and $%
\epsilon _{d}$) have SU(2) symmetry, thus the flavor symmetry (SU(3)$_{\text{%
f}}$, SU(4)$_{\text{f}}$ and SU(5)$_{\text{f}}$) are the natural extensions
of SU(2)$_{\text{f}}$. Then this paper deduces the important baryons, from
the deduced quarks, using the qqq baryon model and a phenomenological
binding energy formula as well as SU(4). At the same time, it also deduces
the important mesons, from the deduced quarks, using the q$\overline{\text{q}%
}$ meson model and a phenomenological binding energy formula. The deduced
masses of the baryons and mesons are 98\% consistent with experimental
results. The deduced quantum numbers of the baryons and mesons match with
the experimental results exactly.{\small \ }

The current quarks u, c and t have Q = +$\frac{2}{3}$; and the current
quarks d, s and b have Q = -$\frac{1}{3}$. This case seems to indicate that
there are two kinds of the quarks. One of them has Q = +$\frac{2}{3}$, and
the other one has Q =\ -$\frac{1}{3}$. The Quark Model implicitly assumes
that all quarks are elementary particles, but a quark with a larger mass
always automatically decays into a smaller mass quark in a very short time,
and the smaller mass quark always decays into the u-quark or d-quark. This
situation might show that the quarks are not all elementary particles. The
quark pairs (u - $\overline{\text{u}}$), (d - $\overline{\text{d}}$), (s - $%
\overline{\text{s}}$), (c - $\overline{\text{c}}$) and (b - $\overline{\text{%
b}}$) can be excited from the vacuum. At the same time, these quark pairs
can annihilate back to the vacuum. These facts might imply that the quarks
are from the vacuum. Thus we infer that there might be only two kinds of
elementary quarks $\epsilon _{u}$ and $\epsilon _{d}$ in the vacuum state
essentially and all quarks (including u and d) are their excited states from
the vacuum state. The quarks u and c are the excited states of $\epsilon
_{u} $; and the quarks d, s and b are the excited states of $\epsilon _{d}$%
.\ The following sections will show that our inferences might be correct.\
Since the mass of the top quark (t) is much larger than other quark masses
(about 185 time proton mass), we cannot deduce the mass of the top quark
using the phenomenological formulae. How to deduce the top mass is an open
problem of this paper.

In order to deduce baryons and mesons from quarks, we have to deduce quarks
first.{\small \ }We are lucky, because there is not any baryon or meson that
contains the top quark, we do not need to deduce the mass of the top quark
(t) in this paper.\ \ {\small \ \ \ \ }

\ \ \ \ \ \ \ \ \ \ \ \ \ \ \ \ \ \ \ \ \ \ \ \ \ \ \ \ \ \ \ \ \ \ \ \ \ \
\ \ \ \ \ \ \ \ \ \ \ \ \ \ \ \ \ \ \ 

\ \ \ \ \ \ \ \ \ \ \ \ \ \ \ \ \ \ \ \ \ \ \ \ \ \ \ \ \ \ \ \ \ \ \ \ \ \
\ \ \ \ \ \ \ \ \ \ \ \ \ \ \ \linebreak \textbf{2.} \textbf{Using
Phenomenological Formulae, Deducing the Masses and Flavors }

\textbf{of Quarks from Two Elementary Quarks }$\epsilon _{u}$\textbf{\ and }$%
\epsilon _{d}$

\ \ \ \ \ \ \ \ \ \ \ \ \ \ \ \ \ \ \ \ \ \ \ \ \ \ \ \ \ \ \ \ \ \ \ \ \ \
\ \ \ \ \ \ \ \ \ \ \ \ \ \ \ \ \ \ \ \ 

In this section, we assume that there are only two kinds of unflavored
elementary quarks $\epsilon _{u}$ and $\epsilon _{d}$ in the vacuum state
(there is not any other kind of quark in the vacuum state). Using
phenomenological formulae, we deduce the masses and quantum numbers of the
current quarks from the two kinds of elementary quarks $\epsilon _{u}$ and $%
\epsilon _{d}$.

\ \ \ \ \ \ \ \ \ \ \ \ \ \ \ \ \ \ \ \ \ \ \ \ \ \ \ \ \ \ \ \ \ \ \
\linebreak \textbf{2.1.} \textbf{The two kinds of elementary quarks }$%
\epsilon _{u}$\textbf{\ and }$\epsilon _{d}$ \textbf{in the vacuum} \textbf{%
state}

\ \ \ \ \ \ \ \ \ \ \ \ \ \ \ \ \ \ \ \ \ \ \ \ \ \ \ \ \ \ \ \ \ \ \ \ \ \
\qquad \qquad

We assume that there is only one kind of unflavored (S = C = B = 0)
elementary quark family $\epsilon $ with a color (or red or blue or yellow),
a baryon number $\mathbb{B}$ = $\frac{1}{3},$ a\ spin s = $\frac{1}{2},$ an
isospin I = $\frac{1}{2}$\ and two isospin states ($\epsilon _{u}$ has I$%
_{z} $ = $\frac{1}{2}$ and Q = +$\frac{2}{3}$, while $\epsilon _{d}$ has I$%
_{z}$ = -$\frac{1}{2}$ and Q = -$\frac{1}{3}$). For $\epsilon _{u}$ (or $%
\epsilon _{d}$),\ there are three different colored (or red or blue or
yellow) members. Thus there are six kinds of elementary quarks in the $%
\epsilon $-elementary quark family. The elementary quarks $\epsilon _{u}$
and $\epsilon _{d}$ have flavor SU(2)$_{\text{f}}$ symmetry.

They are essentially in the vacuum state. When they are in the vacuum state,
their baryon numbers, electric charges, spins, isospins and masses (B = Q =
I = I$_{z}$ = s = m$_{\epsilon _{u}}$ = m$_{\epsilon _{d}}$ = 0) cannot be
seen. Although we cannot see them, as physical vacuum background, they
really exist in the vacuum state. Once they obtain enough energies, they may
be excited from the vacuum state to form observable baryons or mesons.

\ \ \ \ \ \ \ \ \ \ \ \ \ \ \ \ \ \ \ \ \ \ \ \ \ \ \ \ \ \ \ \ \ \ \ \ \ \
\ \ \ \ \ \ \ \ \ \ \ \ \ \ \ \ \ \ \ \ \ \ \ \ \ \ \ \ \ \ \ \ \ \ \ \ \ \
\ \ \ \ \ \ \ \ \ \ \ \ \ \ \ \ \ \ \ \ \ \ \ \ \ \ \ \ \linebreak \textbf{%
2.2. The normally excited quarks u\ and d}

\ \ \ \ \ \ \ \ \ \ \ \ \ \ \ \ \ \ \ \ \ \ \ \ \ \ \ \ \ \ \ \ \ \ \ \ \ \
\ \ \ \ \ \ \ \ \ \ \ \ \ \ \ \ \ \ \qquad \qquad \textbf{\ }\ \ \ \ \ \ \ \
\ \ \ \ \ \ \ \ \ \ 

Generally, when a particle normally excited from the vacuum state, it will
keep its intrinsic quantum numbers unchanged and get a continuous energy
spectrum from the lowest energy (the rest mass) to infinite. As a colored
(or red or blue or yellow) elementary quark $\epsilon _{u}$ is normally
excited from the vacuum state, it will excite into the observable state
inside a hadron with the color (or red or blue or yellow), the electric
charge Q = $\frac{2}{3}$, the spin s = the isospin I = the isospin
z-component = $\frac{1}{2}$ and the S = C = B = 0; and it will get an energy
of a continuous energy spectrum from the lowest energy (the rest mass m$_{u}$%
) to infinite. Comparing its quantum numbers with the current quark quantum
numbers \cite{Quark}, we recognize that it is the current u-quark.
Similarly, as a colored (or red or blue or yellow) elementary quark $%
\epsilon _{d}$ is normally excited from the vacuum state, it will excite
into the observable state inside a hadron with $\mathbb{B}$ = $\frac{1}{3}$,
Q = - $\frac{1}{3}$, I = s = $\frac{1}{2}$, I$_{z}$ = -$\frac{1}{2}$, S = C
= B = 0 and it will get an energy of a continuous energy spectrum from the
lowest energy (the rest mass m$_{d}$) to infinite. Comparing its quantum
numbers with the current quark quantum numbers \cite{Quark}, we recognize
that it is the current d-quark. Since $\epsilon _{u}$ and $\epsilon _{d}$
have the flavor SU(2)$_{\text{f}}$ symmetry, the normally excited quarks u
and d have the flavor SU(2)$_{\text{f}}$ symmetry also. Thus we have 
\begin{equation}
\begin{tabular}{l}
the u-quark: $\mathbb{B}$ =$\frac{1}{3},$ Q = $\frac{2}{3}$, I = I$_{z}$ = s
= $\frac{1}{2}$, S = C = B = 0, m$_{u}$; \\ 
the d-quark: $\mathbb{B}$ = $\frac{1}{3}$, Q = - $\frac{1}{3}$, I = -I$_{z}$%
= s = $\frac{1}{2}$, S = C = B = 0, m$_{d}$.%
\end{tabular}
\label{ud}
\end{equation}%
Let us find the rest masses m$_{u}$ of the u-quark and the rest mass m$_{d}$
of the d-quark now.

According to the Quark Model \cite{Quark Model}, a proton is composed of
three quarks (uud) and a neutron is composed of three quarks (udd) also.
Thus the proton mass M$_{\text{p}}$ and the neutron mass M$_{\text{n}}$
generally will be

\begin{equation}
\begin{tabular}{l}
M$_{\text{p}}$ = m$_{\text{u}}$+m$_{\text{u}}$+m$_{\text{d}}-\left\vert 
\text{E}_{\text{Bind}}\text{(p)}\right\vert $, \\ 
M$_{\text{n}}$ = m$_{\text{u}}$+m$_{\text{d}}$+m$_{\text{d}}-\left\vert 
\text{E}_{\text{Bind}}\text{(n)}\right\vert $,%
\end{tabular}
\label{Mp and Mn}
\end{equation}%
where $\left\vert \text{E}_{\text{Bind}}\text{(p)}\right\vert $ and $%
\left\vert \text{E}_{\text{Bind}}\text{(n)}\right\vert $ are the binding
energy of the three quarks inside p and n. Since the mass difference between
p(938) and n(940) is very small, we omit the differences of the masses and
binding energies between p(938) and n(940). Taking an average mass (939) of
p and n masses, from (\ref{Mp and Mn}), we get

\begin{eqnarray*}
\text{m}_{\text{u}}\text{+m}_{\text{u}}\text{+m}_{\text{d}}\text{-}%
\left\vert \text{E}_{\text{Bind}}\text{(p)}\right\vert &\text{=}&\text{m}_{%
\text{u}}\text{+m}_{\text{d}}\text{+m}_{\text{d}}\text{-}\left\vert \text{E}%
_{\text{Bind}}\text{(n)}\right\vert \text{= 939} \\
\left\vert \text{E}_{\text{Bind}}\text{(p)}\right\vert &\approx &\text{ }%
\left\vert \text{E}_{\text{Bind}}\text{(n)}\right\vert \text{.}
\end{eqnarray*}%
Then we have approximately: 
\begin{equation}
\text{ m}_{\text{u}}\approx \text{m}_{\text{d}}\text{,}  \label{Mm}
\end{equation}%
$\left\vert \text{E}_{\text{Bind}}\text{(p)}\right\vert $ $\approx $ $%
\left\vert \text{E}_{\text{Bind}}\text{(n)}\right\vert $ = $\left\vert \text{%
E}_{\text{Bind}}\right\vert $ is an unknown complex function. As a
phenomenological approximation, we assume that $\left\vert \text{E}_{\text{%
Bind}}\right\vert $ = 3$\Delta $\ ($\Delta $ is an unknown constant).
Because any high energy particle has not destroyed a proton or a neutron to
get a free quark, we assume that $\Delta $ is an unknown huge constant ($%
\Delta $ $\gg $ m$_{\text{P}}$= 938 Mev). From $\left\vert \text{E}_{\text{%
Bind}}\right\vert $ = 3$\Delta $, (\ref{Mp and Mn}) and (\ref{Mm}), we find m%
$_{\text{u}}$= m$_{\text{d}}$ = 313 + $\Delta $:

\begin{equation}
\begin{tabular}{l}
m$_{\text{u}}$= m$_{\text{d}}$ = 313 + $\Delta \text{ }$ \\ 
$\text{\ }\Delta $ = $\frac{1}{3}\left\vert \text{E}_{\text{Bind}%
}\right\vert $ $\gg $ m$_{\text{P}}$=938Mev.%
\end{tabular}
\label{313andDalta}
\end{equation}

Now we have two unflavored quarks u(313+$\Delta $) and d(313+$\Delta $).
They are the most important quarks. They compose the most important baryons
(p and n) and mesons ($\pi ^{\pm }$ and $\pi ^{0}$).\ How do the flavored
quarks (s, c and b) come out?\ 

\ \ \ \ \ \ \ \ \ \ \ \ \ \ \ \ \ \ \ \ \ \ \ \ \ \ \ \ \ \ \ \ \ \ \ \ \ \
\ \ \ \ \ \ \ \ \ \ \ \ \ \ \ \ \ \ \ \ \ \ \ \ \ \ \ \ \ \ \ \ \ \ \ \ \ \
\ \ \ \ \ \ \ \ \ \ \linebreak \textbf{2.3.} \textbf{Using phenomenological
formulae,}\ \textbf{deducing} \textbf{the flavored }

\ \ \textbf{quarks s, c and b from the elementary quarks }$\epsilon _{u}$%
\textbf{\ or }$\epsilon _{d}$\textbf{\ }

\ \ \ \ \ \ \ \ \ \ \ \ \ \ \ \ \ \ \ \ \ \ \ \ \ \ \ \ \ \ \ \ \ \ \ \ \ \
\ \ \ \ \ \ \ \ \ \ \ \ 

The flavored quarks (s, c and b) are also the excited states of the
elementary quarks $\epsilon _{u}$ or $\epsilon _{d}$; they all have the
baryon number $\mathbb{B}$ = $\frac{1}{3}$ and the spin s = the isosin = $%
\frac{1}{2}$ because $\epsilon $ has $\mathbb{B}$ = $\frac{1}{3}$ and s = $%
\frac{1}{2}$. The excited quarks of $\epsilon _{u}$ have the electric charge
Q = +$\frac{2}{3}$ since $\epsilon _{u}$ has Q = +$\frac{2}{3}$. The excited
quarks of $\epsilon _{d}$ have the electric charge Q = -$\frac{1}{3}$ since $%
\epsilon _{d}$ has Q = -$\frac{1}{3}$. Thus the current flavored quark c
with Q = +$\frac{2}{3}$ is the excited quark of $\epsilon _{u}$; the current
flavored quarks s and b with Q = -$\frac{1}{3}$ are the excited quarks of $%
\epsilon _{d}$. Although the flavored quarks are also the excited states of
the elementary quarks, there are large differences from the normally excited
states (u and d) of $\epsilon _{u}$\ and $\epsilon _{d}$:

(1) A flavored quark has a flavor that is different from the noemally
excited quarks u and d that are unflavored quarks.

(2) None of the flavored quarks can compose any stable baryon or meson; this
case means that they are short lifetime quarks.

(3) All flavored quarks are single isospin (I = 0) particles that are
different from the unflavored quarks u and d with I = $\frac{1}{2}$.

(4) They have quantized energies (masses) that are different from the
continuous energy spectrum of u and d.

In order to deduce the quantizing energies (masses) of the\ flavored quarks,
let us recall Planck's and Bohr's quantizations. Using the quantization
condition ($\varepsilon $ = nh$\nu $ \ n = 0, 1, 2, ...), Planck \cite%
{Planck} selected correct energies from a classic continuous energy
spectrum. Using the quantization condition (L = $\frac{nh}{2\pi }$, \ n = 1,
2, 3, ... ), Bohr \cite{Bohr} selects correct orbits (energy levels) from
the classic infinite orbits (continuous energy spectrum). Similarly,
extending the Planck-Bohr quantization from the linear funcrion of n to a
quadric function of n, we get a quantization condition (a phenomenological
quark mass spectrum which depends on the electric charge of the possible
flavored quark). If an excited state with a quantized energy E$_{\text{Q }}$%
(mass): 
\begin{equation}
\text{E}_{\text{Q\ }}\text{= m}_{0}\text{+360(1+}\widetilde{\text{Q}}\text{%
)[(2n+}\frac{1}{2}\widetilde{\text{Q}}\text{)}^{2}\text{-}\frac{1}{4}\text{%
Sign(Q)n],\ \ \ n+}\widetilde{\text{Q}}\text{= 1, 2, 3, ... ,}
\label{Qk-Mass}
\end{equation}%
it is a possible flavored quark with a flavored value : 
\begin{equation}
\text{Flavor value = Sign(Q)}\times \text{1, }\ \text{n+}\widetilde{\text{Q}}%
\text{= 1, 2, 3, ... .\ }  \label{flavor}
\end{equation}%
Where Q is the electric charge of the excited state; m$_{0}$ = m$_{u}$= m$%
_{d}$ = (313+$\Delta $) (\ref{Qk-Mass}); $\widetilde{\text{Q}}$ $\equiv
\left\vert \text{Q-}\frac{2}{3}\right\vert $, for Q = +$\frac{2}{3}$, $%
\widetilde{\text{Q}}$ = 0 , for Q = -$\frac{1}{3}$, $\widetilde{\text{Q}}$ =
1. The Sign(Q) is signum of the Q; for the flavored excited states of $%
\epsilon _{u}$, Q = +$\frac{2}{3}$\ and Sign(Q) = \textquotedblleft\
+\textquotedblright ; for the flavored excited state of $\epsilon _{d}$, Q =
-$\frac{1}{3}$ and Sign(Q) = \textquotedblleft -\textquotedblright . This
phenomenological mass formula has only one arbitrary parameter m$_{0}$ that
is determined by proton mass and newtron mass (\ref{313andDalta}).

From (\ref{Qk-Mass}) and (\ref{flavor}), for the excited states of $\epsilon
_{u}$, Q = +$\frac{2}{3}$, $\widetilde{\text{Q}}$ = 0, Sign(Q)= + 1 and n =
1, 2, 3, ...; for the excited states of $\epsilon _{d}$, Q = -$\frac{1}{3}$, 
$\widetilde{\text{Q}}$ = 1, Sign(Q) = - 1 and n = 0, 1, 2, ... . Thus we
have the masses and flavor values of the possible flavored excited quarks of 
$\epsilon _{u}$ and $\epsilon _{d}$: 
\begin{equation}
\begin{tabular}{l}
for $\epsilon _{u}$, m$_{_{+\frac{2}{3}}\text{\ }}$= m$_{0}$+360[(2n)$^{2}$-$%
\frac{1}{4}$n] \ \ and\ flavor = +1\ \ \ \ \ \ n = 1, 2, 3, ... ; \\ 
for $\epsilon _{d}$, m$_{-\frac{1}{3}}$= m$_{0}$+720[(2n+$\frac{1}{2}$)$^{2}$%
+$\frac{1}{4}$n] \ \ and\ flavor = -1\ \ \ n = 0, 1, 2, ... .%
\end{tabular}
\label{Q-M}
\end{equation}%
Putting the m$_{0}$ = 313 + $\Delta $ and n values into (\ref{Q-M}), we find
the masses and flavor values of the flavored excited states (possible
flavored quarks) shown in Table 1. We have already deduced two unflavored
quarks u and d (\ref{ud}) and (\ref{313andDalta}) shown in Table 1 also:\ 

\ \ \ \ \ \ \ \ \ \ \ \ \ \ \ \ \ \ \ \ \ \ \ \ \ \ \ \ \ \ \ \ \ \ \ \ \ \
\ \ \ \ \ \ \ \ \ \ \ \ \ \ \ \ \ \ \ \ \ \ \ \ \ 

\ 
\begin{tabular}{l}
\ Table 1: \ The masses and flavors of the excited states (possible quarks)
\\ 
\ 
\begin{tabular}{|l|l|l|l|l|}
\hline
\ \ \ \ \ \ \ \ n & \ \ \ \ \ \ 0 & \ \ \ \ \ \ \ 1 & \ \ \ \ \ \ \ 2 & \ \
\ \ \ \ \ \ 3 \\ \hline
\begin{tabular}{l}
{\small Excited} \\ 
{\small states of }$\epsilon _{u}$%
\end{tabular}
& 
\begin{tabular}{l}
{\small u(}$\Delta ${\small +313)} \\ 
{\small S=C=B=0}%
\end{tabular}
& 
\begin{tabular}{l}
{\small c(}$\Delta ${\small +1663)} \\ 
{\small C = 1}%
\end{tabular}
& 
\begin{tabular}{l}
{\small c}$^{\ast }(\Delta ${\small +5893)} \\ 
{\small C}$^{\ast }${\small \ = 1}%
\end{tabular}
& 
\begin{tabular}{l}
\ {\small c}$^{2\ast }${\small (}$\Delta ${\small +13003)} \\ 
{\small C}$^{2\ast }${\small \ = 1}%
\end{tabular}
\\ \hline
{*}******** & ******** & ********* & ********** & ************ \\ \hline
\ \ \ \ \ \ \ \ n & \ \ \ \ \ -1 & \ \ \ \ \ \ \ 0 & \ \ \ \ \ \ \ 1 & \ \ \
\ \ \ \ \ 2 \\ \hline
\begin{tabular}{l}
{\small xcited} \\ 
{\small states of }$\epsilon _{d}$%
\end{tabular}
& 
\begin{tabular}{l}
{\small d(}$\Delta ${\small +313)} \\ 
{\small S=C=B=0}%
\end{tabular}
& 
\begin{tabular}{l}
{\small s(}$\Delta ${\small +493)} \\ 
{\small S = -1}%
\end{tabular}
& 
\begin{tabular}{l}
{\small b(}$\Delta ${\small +4993)} \\ 
{\small B = -1}%
\end{tabular}
& 
\begin{tabular}{l}
{\small b}$^{\ast }${\small (}$\Delta ${\small +15253)} \\ 
{\small B}$^{\ast }${\small \ = -1.}%
\end{tabular}
\\ \hline
\end{tabular}
\\ 
\ {\small In Table 1, we use the quark names to show their quantum numbers
as usual. }%
\end{tabular}

\ \ \ \ \ \ \ \ \ \ \ \ \ \ \ \ \ \ \ \ \ \ \ \ \ \ \ \ \ \ \ \ \ 

Table 1 shows that there are three kinds of possible quarks: the first kind
for n \TEXTsymbol{<} 1 light quarks u(313+$\Delta $), d(313+$\Delta $) and
s(493+$\Delta $); the second kind for n =1, heavy quarks c(1663+$\Delta $)
and b(4993+$\Delta $); the third kind for n \TEXTsymbol{>} 1 the flavored
excited states {\small c}$^{\ast }$(5893+$\Delta $), {\small c}$^{2\ast }$%
(13003+$\Delta $) and b$^{\ast }$(15253+$\Delta $). Today's experiments show
that for n \TEXTsymbol{<} 1 light quarks, we have found almost all possible
baryons and mesons made by the light quarks; for n = 1, the heavy quarks
c(1663+$\Delta $) and b(4993+$\Delta $), we have found many baryons and
mesons made by the heavy quarks; for n \TEXTsymbol{>} 1, the higher energy
flavored excited states {\small c}$^{\ast }$(5893+$\Delta $), {\small c}$%
^{2\ast }$(13003+$\Delta $) and b$^{\ast }$(15253+$\Delta $), there is not
any confirmed baryon or meson that is composed by these excited states. From
today's experimental result, we cannot think these excited states are quarks
that can compose baryons and mesons. Thus, we conclude that the ground
unflavored excited state u(313+$\Delta $) with Q = +$\frac{2}{3}$, the
ground unflavored excited state d(313+$\Delta $) with Q = - $\frac{1}{3}$,
the ground excited strange state s(493+$\Delta $) with S = -1, the ground
charmed excited state c(1663+$\Delta $) with C = +1 and the ground bottom
excited state b(4993+$\Delta $) with B = -1 are the quarks that can compose
baryons and mesons, shown in Table 2; but the flavored excited states 
{\small c}$^{\ast }$(5893+$\Delta $), {\small c}$^{2\ast }$(13003+$\Delta $)
and b$^{\ast }$(15253+$\Delta $) are not quarks. The foundation of this
conclusion is on today's experimental results. As new experimental
technology and equipment become available, in future, many new baryons and
mesons might be discovered. Some new quark might be discovered also.

\ \ \ \ \ \ \ \ \ \ \ \ \ \ \ \ \ \ \ \ \ \ \ \ \ \ \ \ \ \ \ \ \ \ \ \ \ \
\ \ \ \ \ \ \ \ \ \ \ \ \ \ \ \ \ \ \ \ \ \ \ \ \ \ \ \ \ \ \ \ \ \ 

\begin{tabular}{l}
\ \ \ \ \ \ Table 2: \ The quarks and their quantum numbers and masses \\ 
\begin{tabular}{|l|l|l|l|l|l|}
\hline
Quark & u(313) & d(313) & s(493) & c(1663) & b(4993) \\ \hline
$\mathbb{B}$ - Baryon number & $\frac{{\small 1}}{3}$ & $\frac{{\small 1}}{3}
$ & $\frac{{\small 1}}{3}$ & $\frac{{\small 1}}{3}$ & $\frac{{\small 1}}{3}$
\\ \hline
s - Spin & $\frac{{\small 1}}{2}$ & $\frac{{\small 1}}{2}$ & $\frac{{\small 1%
}}{2}$ & $\frac{{\small 1}}{2}$ & $\frac{{\small 1}}{2}$ \\ \hline
I - Isospin \ \ \ \  & $\frac{{\small 1}}{2}$ & $\frac{{\small 1}}{2}$ & 0 & 
0 & 0 \\ \hline
I$_{z}$ - Isospin z-component & $\frac{{\small 1}}{2}$ & -$\frac{{\small 1}}{%
2}$ & 0 & 0 & 0 \\ \hline
S - Strangeness & 0 & 0 & -1 & 0 & 0 \\ \hline
C - Charm & 0 & 0 & 0 & +1 & 0 \\ \hline
B - Bottomness & 0 & 0 & 0 & 0 & -1 \\ \hline
Q - Electric Charge & +$\frac{{\small 2}}{3}$ & -$\frac{{\small 1}}{3}$ & -$%
\frac{{\small 1}}{3}$ & +$\frac{{\small 2}}{3}$ & -$\frac{{\small 1}}{3}$ \\ 
\hline
m - mass (Mev) & 313+$\Delta $ & 313+$\Delta $ & 493+$\Delta $ & 1663+$%
\Delta $ & 4993+$\Delta $ \\ \hline
\end{tabular}
\\ 
{\small u(313) and d(313) are the two component states I}$_{z}${\small \ = }$%
\frac{1}{2}${\small (u(313)) and I}$_{z}${\small \ = -}$\frac{1}{2}${\small %
(d(313))} \\ 
{\small \ of one isospin quark q}$_{\text{{\small N}}}${\small (313) with I
= }$\frac{1}{2}$.%
\end{tabular}

\ \ \ \ \ \ \ \ \ \ \ \ \ \ \ \ \ \ \ \ \ \ \ \ \ \ \ \ \ \ 

When the quark pairs (u - $\overline{\text{u}}$), (d - $\overline{\text{d}}$%
), (s - $\overline{\text{s}}$), (c - $\overline{\text{c}}$) and (b - $%
\overline{\text{b}}$) annihilate back to the vacuum state, the u-quark and
the c-quark will decay back to the elementary quark $\epsilon _{u}$ with
unobservable color, electric charge, spin, isospin, mass and S = C = B = 0
in the vacuum state; and the d-quark, the s-quark and the b-quark will decay
back to the elementary quark $\epsilon _{d}$ with unobservable color,
electric charge, spin, isospin, mass and S = C = B = 0 in the vacuum state.

\ \ \ \ \ \ \ \ \ \ \ \ \ \ \ \ \ \ \ \ \ \ \ \ \ \ \ \ \ \ \ \ \linebreak 
\textbf{2.4} \textbf{The extensions of SU(3)}$_{\text{f}}$\textbf{, SU(4)}$_{%
\text{f}}$\textbf{\ and SU(5)}$_{\text{f}}$\textbf{\ from SU(2)}$_{\text{f}}$%
\ \ 

\ \ \ \ \ \ \ \ \ \ \ \ \ \ \ \ \ \ \ \ \ \ \ \ \ \ \ \ \ \ \ \ \ \ \ \ \ \
\ \ \ \ \ \ \ \ \ \ \ \ \ \ \qquad \qquad\ \ \ \ \ \ \ \ \ \ \ \ \ \ \ \ \ \
\ \ \ \ \ \ \ \ \ 

Since the elementary quarks $\epsilon _{u}$ and $\epsilon _{d}$ have the
flavor SU(2)$_{\text{f}}$ symmetry, their normally excited quarks (u and d)
have the flavor SU(2)$_{\text{f}}$ symmetry also. Because the five quarks
(u, d, s, c and b) are all the excited states of the elementary quarks $%
\epsilon _{u}$ or $\epsilon _{d}$, SU(3)$_{\text{f}}$ (basic quarks u, d and
s), SU(4)$_{\text{f}}$ (basic quarks u, d, s and c) and SU(5)$_{\text{f}}$
(basic quarks u, d, s, c and b) are the natural extensions of SU(2)$_{\text{f%
}}$. Since $\Delta $ $\gg $ m$_{\text{p}}$ = 938 Mev, from Table 2, m$_{u}$
(=$\Delta $+313) = m$_{d\text{ }}$(=$\Delta $+313) $\approx $\ m$_{s}$ (= $%
\Delta $+493) $\approx $ m$_{c}$ (= $\Delta $+1663) $\approx $ \ m$_{b}$ (= $%
\Delta $+4993). Thus SU(4)$_{\text{f}}$, and SU(5)$_{\text{f}}$ are not
badly broken by the quark masses. We will use SU(4) and the qqq baryon model
to deduce baryons from the deduced quarks in Table 2.

\ \ \ \ \ \ \ \ \ \ \ \ \ \ \ \ \ \ \ \ \ \ \ \ \ \ \ \ \ \ \ \ \ \ \ \ \ \
\ \ \ \ \ \ \ \ \ 

\ \ \ \ \ \ \ \ \ \ \ \ \ \ \ \ \ \ \ \ \ \ \ \ \ \ \ \ \ \ \ \ \ \ \ \ \ \
\ \ \ \ \ \ \ \ \ \ \ \ \ \ \ \ \textbf{\ }\linebreak \textbf{3.} \ \ 
\textbf{Using the Sum Laws, SU(4) and a Phenomenological Binding Energy}

\ \ \textbf{Formula, Deducing} \textbf{the Important Baryons from the
Deduced Quarks}

\ \ \ \ \ \ \ \ \ \ \ \ \ \ \ \ \ \ \ \ \ \ \ \ \ \ \ \ \ \ \ \ \ \ \ \ \ \
\ \ \ \ \ \ \ \ \ \ \ \ 

According to the Quark Model, a colorless baryon is composed of three
different\ colored quarks. From Table 2, we can see that there is a term $%
\Delta \ $inside the masses of each quark. $\Delta $ is an unknown huge
constant. Since the masses of the three quarks in a baryon are huge (from $%
\Delta $) and the mass of the baryon composed by the three quarks is not, we
infer that there will be a part of binding energy (E$_{Bind}$ = - 3$\Delta $%
) to cancel the 3$\Delta $ of the three quark masses. Thus the baryon mass
will essentially be: 
\begin{equation}
\begin{tabular}{l}
$M_{\text{B}}\text{ }\text{= m}_{q_{1}}^{\ast }\text{{\small +\ m}}%
_{q_{2}}^{\ast }\text{{\small \ +m}}_{q_{3}}^{\ast }\text{{\small -}}%
\left\vert \text{E}_{Bind}\right\vert $ \\ 
$\text{= }\text{(m}_{q_{1}}\text{+}\Delta \text{)+(m}_{q_{2}}\text{+}\Delta 
\text{)+(m}_{q_{3}}\text{+}\Delta \text{)-3}\Delta $ \\ 
$\text{= }\text{m}_{q_{1}}\text{ + m}_{q_{2}}\text{ + m}_{q_{3}}\equiv \text{
M}_{\text{123}}\text{.}$%
\end{tabular}
\label{B-MASS}
\end{equation}%
Therefore we will omit the term $\Delta \ $in the three quark masses of the
baryon and the term ($-3\Delta )\ $in the binding energy of the baryon when
we deduce the masses of baryons. The sum laws of the baryon number $\mathbb{B%
}$, the strange number S, the charmed number C, the bottom number B, the
electric charge Q and the isospin z-component I$_{z}$ of a baryon are:

\begin{equation}
\begin{tabular}{ll}
$\mathbb{B}_{\text{Baryon}}\text{ }\text{= }\mathbb{B}_{q_{1}}\text{+ }%
\mathbb{B}_{q_{2}}\text{+ }\mathbb{B}_{q_{3}}\text{, }$ & $\text{S}_{\text{%
Baryon}}\text{ }\text{= S}_{q_{1}}\text{+ S}_{q_{2}}\text{+ S}_{q_{3}}\text{,%
}$ \\ 
$\text{C}_{\text{Baryon}}\text{ }\text{= C}_{q_{1}}\text{ + C}_{q_{2}}\text{
+ C}_{q_{3}}\text{, }$ & B$_{\text{Baryon}}\text{ =}\text{ B}_{q_{1}}\text{+
B}_{q_{2}}\text{+ B}_{q_{3}}\text{,}$ \\ 
$\text{Q}_{\text{Baryon}}\text{ = Q}_{q_{1}}\text{+ Q}_{q_{2}}\text{+ Q}%
_{_{q_{3}}}$, & I$_{\text{z,Baryon}}$ = I$_{z,q_{1}}$+\ I$_{z,q_{2}}$+ I$%
_{z,q_{3}}$.%
\end{tabular}
\label{SumB}
\end{equation}

\ \ \ \ \ \ \ \ \ \ \ \ \ \ \ \ \ \ \ \ \ \ \ \ \ \ \ \ \ \ \ \ \ \ \ \ \ \
\ \ \ \ \ \ \ \ \ \ \ \ \ \ \ \ \ \ \ \ \ \ \ \ \ \ \ \ \ \ \ \ \ \ \ \ \ \
\ \ \ \ \ \ \ \ \ \ \ \ \ \ \ \ \ \ \ \ \ \ \ \ \ \ \ \ \ \ \ \ \ \ \ \ \ \
\ \ \ \ \ \linebreak \textbf{3.1. Using the sum laws, Deducing the most
important baryons }

\ \ \ \textbf{from the Deduced quarks }

\ \ \ \ \ \ \ \ \ \ \ \ \ \ \ \ \ \ \ \ \ \ \ \ \ \ \ \ \ \ \ \ \ \ \ \ \ \
\ \ \ \ \ \ \ \ \ \ \ \ \ \ \ \ \ \ \ \ \ \ \ \ \ \ \qquad \qquad\ \ \ \ \ \
\ \ \ \ \ \ \ \ \ \ \ \ \ \ \ \ \ \ $\qquad \qquad $\ \ \ \ \ \ \ \ \ \ \ \
\ \ \ \ \ \ \ \ \ \ \ \ \ \ \ \ \ \ \ \ \ \ \ \ \ \ \ \ 

Using the quantum number sum laws (\ref{SumB}) and the mass sum law (\ref%
{B-MASS}), from the S, C, B, Q, I$_{z}$ and the masses of the deduced quarks
in Table 2, omitting the $\Delta $ in the quark masses and the binding
energy (-3$\Delta $), we deduce the most important baryons, shown in Table 3:

\ \ \ \ \ \ \ \ \ \ \ \ \ \ \ \ \ \ \ \ \ \ \ \ \ \ \ \ \ \ \ \ \ \ \ \ \ \
\ \ \ \ \ \ \ \ \ \ \ \ \ \ \ \ \ \ 

\begin{tabular}{l}
\ \ \ Table 3: \ The most important baryons (J$^{\text{P}}$= $\frac{1}{2}%
^{+} $ and $\mathbb{B}$ = $1$) \\ 
\begin{tabular}{|l|l|l|l|l|l|l|}
\hline
Quark$_{1}$ & u(313) & d(313) & s(493) & c(1663) & b(4993) & c$^{\ast }$%
(5893) \\ \hline
Quark$_{2}$ & u(313) & u(313) & u(313) & u(313) & u(313) & u(313) \\ \hline
Quark$_{3}$ & d(313) & d(313) & d(313) & d(313) & d(313) & d(313) \\ \hline
Deduced & p(939) & n(939) & $\Lambda $(1119) & $\Lambda _{c}$(2289) & $%
\Lambda _{b}$(5619) & $\Lambda _{c^{\ast }}^{\ast }$(6519) \\ \hline
Exper. & p(938) & n(940) & $\Lambda $(1116) & $\Lambda _{c}$(2285) & $%
\Lambda _{b}$(5624) & \ \ \ \ \ ? \\ \hline
$\frac{\Delta M}{M}$\% & 0.1 & 0.1 & 0.3 & 0.2 & 0.09 & \ \ \ \ \ ? \\ \hline
\end{tabular}
\\ 
{\small \ }$\Lambda _{c^{\ast }}^{\ast }$(6519) is not a normal baryons; it
might be a quasi-baryon.%
\end{tabular}

\ \ \ \ \ \ \ \ \ \ \ \ \ \ \ \ \ \ \ \ \ \ \ \ \ \ \ \ \ \ \ \ \ \ \ \ \ \
\ \ \ \ \ \ \ \ \ \ \ \ \ \ \ \ \ \ \ \ \ \ \ 

From Table 3, we see that the deduced quantum numbers of the most important
baryons match the experimental results \cite{Baryon} exactly (using the same
names to show the same quantum numbers between the deduced baryons and the
experimental baryons); and the deduced masses of the most important baryons
(p, n, $\Lambda $, $\Lambda _{c}$ and $\Lambda _{b}$) are 99.7\% consistent
with experimental results. The baryon masses are deduced only using sum law M%
$_{\text{B}}\text{ = }\text{m}_{q_{1}}\text{ + m}_{q_{2}}\text{ + m}_{q_{3}}%
\text{. }$These cases might show that the deduced masses of the quarks could
indeed be correct. $\Lambda _{c^{\ast }}^{\ast }$(6519) is not a normal
baryon, it might be a quasi-baryon.

The mass of a baryon essentially is the sum $\text{M}_{\text{123}}$ (\ref%
{B-MASS}) of the three quark masses inside the baryon. For higher isospin
(I) and higher spin (J) baryons, adding a small phenomenological adjustment
binding energy ($\Delta $e), we can find the masses of the baryons with
higher I and J:%
\begin{equation}
\begin{tabular}{l}
M$_{\text{Baryon}}$ = \ $\text{m}_{q_{1}}\text{+ m}_{q_{2}}\text{+ m}%
_{q_{3}} $+ $\Delta e$ = $\text{M}_{\text{123}}$ + $\Delta e$, \\ 
$\Delta $e = 68 [$\Delta $I + 3$\Delta $J + 2C$\times $I$\times \delta $($%
\Delta $J)].%
\end{tabular}
\label{Ebin of B}
\end{equation}%
Where $\Delta $I is the difference between the isospin of the baryon and the
minimum isospin of the three quark system in the baryon,\ $\Delta $J is the
difference between the spin J of the baryon and the minimum J of the three
quark system in the baryon (see Table 4 and Table 9). C is the charm number
and \ I is the isospin number of the baryon. $\delta $($\Delta $J) is a
Dirac $\delta $ function. For $\Delta $J = 0, $\delta $($\Delta $J) = 1; for 
$\Delta $J $\neq $ 0, $\delta $($\Delta $J) = 0.

Using the sum laws (\ref{SumB}), the mass formula (\ref{Ebin of B}), a
flavor-spin SU(6) and the qqq baryon model of the Quark Model, we can deduce
the masses and the quantum numbers of the baryons from the masses and
quantum numbers of the deduced quarks in Table 2.

\ \ \ \ \ \ \ \ \ \ \ \ \ \ \ \ \ \ \ \ \ \ \ \ \ \ \ \ \ \ \ \ \ \ \ \ \ \
\ \linebreak \textbf{3.2. Deducing the baryons of Octet and Decuplet using
the sum laws and }

\ \ \ \textbf{a phenomenological binding energy formula from the deduced
quarks}

\ \ \ \ \ \ \ \ \ \ \ \ \ \ \ \ \ \ \ \ \ \ \ \ \ \ \ \ \ \ \ \ \ \ \ \ \ \
\ \ \ \ \ \ \ \ \ \ \ \ \ \ \ \ \textbf{\ }\ \ \ \ \ \ \ \ \ \qquad \qquad\
\ \ \ \ \ \ \ \ \ \ \ \ \ \ \ 

The flavor and spin of the \textquotedblleft ordinary\textquotedblright\
quarks (u, d and s) may be combined\ in an approximate flavor-spin SU(6)\ in
which the six basic states are u$\uparrow $, u$\downarrow $,\ d$\uparrow $, d%
$\downarrow $, s$\uparrow $ and s$\downarrow $ ($\uparrow $, $\downarrow $ =
spin up, down). Then the baryons belong to the multiplets on the right side
of \ 

\begin{equation}
\text{\textbf{6} }\mathbf{\otimes }\text{ \textbf{6} }\mathbf{\otimes }\text{
\textbf{6} = \textbf{56}}_{S}\text{\ }\mathbf{\oplus }\text{ \textbf{70}}_{M}%
\text{ }\mathbf{\oplus }\text{ \textbf{70}}_{M}\text{ }\mathbf{\oplus }\text{
\textbf{20}}_{A}\text{.}  \label{6x6x6}
\end{equation}%
These SU(6) multiplets decompose into flavor SU(3)$_{\text{f}}$ \cite{SU(3)}
multiplets as follows:

\begin{equation}
\begin{tabular}{l}
$\text{\textbf{56}}_{S}$ = $^{4}$\textbf{10} $\oplus $ $^{2}$\textbf{8,} \\ 
$\text{\textbf{70}}_{M}$ = $^{2}$\textbf{10} $\oplus $ $^{4}$\textbf{8} $%
\oplus $ $^{2}$\textbf{8} $\oplus $ $^{2}$\textbf{1,} \\ 
$\text{\textbf{20}}_{A}$ = $^{2}$\textbf{8} $\oplus $ $^{4}$\textbf{1},%
\end{tabular}
\label{56,70,20}
\end{equation}%
where the superscript (2J + 1) gives the net spin J of the baryons for each
baryon in the SU(3) multiplets. J$^{\text{P}}$ = $\frac{1}{2}^{+}$Octet and J%
$^{\text{P}}$ = $\frac{3}{2}^{+}$ Decuplet together make up the
\textquotedblleft ground-state\textquotedblright\ 56-plet in which the
orbital angular momenta between the quark pairs are zero (so that the
spatial part of the state function is trivially symmetric). \textbf{70}$_{M}$
and\ \textbf{20}$_{A}$ require some excitation of the spacial parts of the
state function. For simplicity we only discuss the \textquotedblleft
ground-state\textquotedblright\ 56-plet in this paper. For the SU(6)
multiplets, since there is not any charmed quark (C = 0), the formula (\ref%
{Ebin of B}) is simplified into:%
\begin{equation}
\begin{tabular}{l}
M$_{\text{Baryon}}$ = $\text{M}_{\text{123}}$ + $\Delta e$, \\ 
$\Delta $e = 68 ($\Delta $I + 3$\Delta $J).%
\end{tabular}
\label{56-plet}
\end{equation}

\ \ \ \ \ \ \ \ \ \ \ \ \ \ \ \ \ \ \ \ \ \ \ \ \ \ \ \ \ \ \ \ \ \ \ \ \ \
\ \ \ \ \ \ \ \ \ \ \ \ \ \ \ \ \ \ \ \ \ \ \ \ \linebreak 3.2.1. Deducing
the baryons of the SU(3) Octet from the deduced quarks

\ \ \ \ \ \ \ \ \ \ \ \ \ \ \ \ \ \ \ \ \ \ \ \ \ \ \ \ \ \ \ \ \ \ \ \ \ \
\ \ \ \ \ \ \ \ \ \ \ \ \ \ \ \ \ \ \ \ 

The baryons p(938), n(940), $\Lambda ^{0}$(1116), $\Sigma $(1193), $\Xi $%
(1318) belong to SU(3)$_{\text{f}}$ Octet with J$^{\text{P}}$ = $\frac{1}{2}%
^{+}$. The SU(3) Octet has given the three quarks for each baryon of the
Octet. Using sum laws (\ref{SumB}) and mass formula (\ref{56-plet}), from
the deduced quarks in Table 2, we deduce the strange number S, the isospin
I, the I$_{z}$, the charge Q and the mass of the baryons shown in Table 4:\
\ 

\ \ \ \ \ \ \ \ \ \ \ \ \ \ \ \ \ \ \ \ \ \ \ \ \ \ \ \ \ \ \ \ \ \ \ \ \ \
\ \ \ \ \ \ \ \ \ \ \ \ \ \ \ \ \ \ \ \ \ \ \ \ \ \ \ \ \ \ \ \ \ \ \ \ \ \
\ \ \ \ \ \ \ \ \ \ \ \ \ \ \ \ \ \ \ \ \ \ \ \ \ \ \ \ \ \ \ \ \ \ \ \ \ \
\ \ \ \ \ \ \ \ \ \ \ \ \ \ \ \ \ \ \ \ \ \ \ \ \ \ \ \ \ \ \ \ \ \ \ \ \ \
\ \ \ \ \ \ \ \ \ \ \ \ \ \ \ \ \ \ \ \ \ \ \ \ \ \ \ \ \ \ \ \ \ \ \ \ \ \
\ \ \ \ \ \ \ \ \ \ \ \ \ \ \ \ \ \ \ \ \ \ \ \ \ \ \ \ \ \ \ \ \ \ \ \ \ \
\ \ \ \ \ \ \ \ \ \ \ \ \ \ \ \ \ \ \ \qquad\ \ \ \ \ \ \ \ \ \ \ \ \ \ \ \
\ \ \ \ \ \ \ \ \ \ \ \ \ \ \ \ \ \ \ \ \ \ \ \ \ \ \ \ \ \ \ \ \ \ \ \ \ \
\ \ \ \ \ \ \ \ \ \ \ \ \qquad\ \ \ \ \ \ 

\begin{tabular}{l}
\ \ \ \ \ \ \ \ \ Table 4: SU(3) \ Octet baryons with \ B = C = 0 and J$^{%
\text{P}}$ = $\frac{1}{2}^{+}$ \\ 
\begin{tabular}{|l|l|l|l|l|l|l|l|l|l|l|l|}
\hline
{\small q}$_{1}$ & {\small q}$_{2}$ & {\small q}$_{3}$ & {\small S} & 
{\small I} & {\small I}$_{z}$ & $\Delta ${\small I} & $\Delta ${\small e} & 
{\small M}$_{123}$ & {\small Deduced} & {\small Exper.} & $\frac{\Delta 
\text{M}}{\text{M}}{\small \%}$ \\ \hline
{\small u(313)} & {\small u} & {\small d} & {\small 0} & $\frac{1}{2}$ & $%
\frac{1}{2}$ & 0 & 0 & {\small 939} & {\small p(939)} & {\small p(938)} & 
{\small 0.1\%} \\ \hline
{\small d(313)} & {\small u} & {\small d} & {\small 0} & $\frac{1}{2}$ & 
{\small -}$\frac{1}{2}$ & 0 & 0 & {\small 939} & {\small n(939)} & {\small %
n(940)} & {\small 0.1\%} \\ \hline
{\small s(493)} & {\small u} & {\small d} & {\small -1} & {\small 0} & 
{\small 0} & 0 & 0 & {\small 1119} & $\Lambda ^{0}${\small (1119)} & $%
\Lambda ^{0}${\small (1116)} & {\small 0.3\%} \\ \hline
{\small s(493)} & {\small u} & {\small u} & {\small -1} & {\small 1} & 
{\small 1} & 1 & 68 & {\small 1187} & $\Sigma ^{+}${\small (1187)} & $\Sigma
^{+}${\small (1189)} & {\small 0.2\%} \\ \hline
{\small s(493)} & {\small u} & {\small d} & {\small -1} & {\small 1} & 
{\small 0} & 1 & 68 & {\small 1187} & $\Sigma ^{0}${\small (1187)} & $\Sigma
^{0}${\small (1193)} & {\small 0.5\% \ } \\ \hline
{\small s(493)} & {\small d} & {\small d} & {\small -1} & {\small 1} & 
{\small -1} & 1 & 68 & {\small 1187} & $\Sigma ^{-}${\small (1187)} & $%
\Sigma ^{-}${\small (1197)} & {\small 0.8\%} \\ \hline
{\small s(493)} & {\small s} & {\small u} & {\small -2} & $\frac{1}{2}$ & $%
\frac{1}{2}$ & 0 & 0 & {\small 1299} & $\Xi ^{0}${\small (1299)} & $\Xi ^{0}$%
{\small (1315)} & {\small 1.2\%} \\ \hline
{\small s(493)} & {\small s} & {\small d} & {\small -2} & $\frac{1}{2}$ & 
{\small -}$\frac{1}{2}$ & 0 & 0 & {\small 1299} & $\Xi ^{-}${\small (1299)}
& $\Xi ^{-}${\small (1321)} & {\small 1.7\%} \\ \hline
\end{tabular}%
\end{tabular}

\ \ \ \ \ \ \ \ \ \ \ \ \ \ \ \ \ \ \ \ \ \ \ \ \ \ \ \ \ \ \ \ \ \ \ \ \ \
\ \ \ \ \ \ \ \ \ \ \ \ \ \ \ \ \ \ \ \ \ \ \ \ \ \ \ \ \ \ \ \ \ \ \ \ \ \
\ \ \ \ \ \ \ \ \ \ \ \ \ \ \ \linebreak 3.2.2. Deducing the baryons of
SU(3) Decuplet from the deduced quarks

\ \ \ \ \ \ \ \ \ \ \ \ \ \ \ \ \ \ \ \ \ \ \ \ \ \ \ \ \ \ \ \ \ \ \ \ \ \
\ \ \ \ \ \ \ \ \ \ \ \ \ \ \ \ \ \ \ \ \ \ \ \ \ \ 

The baryons $\Delta $(1232), $\Sigma $(1385), $\Xi $(1532) and $\Omega ^{-}($%
1672) belong to SU(3) Decuplet with J$^{\text{P}}$ \ = $\frac{3}{2}^{+}$.
The SU(3) Decuplet has given the three quarks for each baryon of the
Decuplet. Using the sum laws (\ref{SumB}) and the mass formula (\ref{56-plet}%
), from the deduced quarks in Table 2, we deduce the strange number S, the
isospin I, the z component I$_{z}$ of the isospin, the charge Q and the
masses of the baryons shown in Table 5 :

\ \ \ \ \ \ \ \ \ \ \ \ \ \ \ \ \ \ \ \ \ \ \ \ \ \ \ \ \ \ \ \ \ \ \ \ \ \
\ \ \ \ \ \ \ \ \ \ \ \ \ \ 

\begin{tabular}{l}
\ \ \ \ \ \ Table 5:\ \ SU(3) Decuplet\ \ with B =\ C = 0\ and\ J$^{\text{P}%
} $ \ = $\frac{3}{2}^{+}$ ($\Delta $J = 1) \\ 
\begin{tabular}{|l|l|l|l|l|l|l|l|l|l|l|l|}
\hline
{\small q}$_{1}$ & {\small q}$_{2}$ & {\small q}$_{3}$ & {\small S} & 
{\small I} & {\small I}$_{z}$ & {\small M}$_{123}$ & $\Delta ${\small I} & $%
\Delta ${\small e} & {\small Deduced} & {\small Exper.} & $\frac{\Delta 
\text{M}}{\text{M}}{\small \%}$ \\ \hline
{\small u(313)} & {\small u} & {\small u} & {\small 0} & $\frac{3}{2}$ & $%
\frac{3}{2}$ & {\small 939} & {\small 1} & {\small 272} & $\Delta ^{++}$%
{\small (1211)} & $\Delta ^{++}${\small (1232)} & {\small 1.7\%} \\ \hline
{\small u(313)} & {\small u} & {\small d} & {\small 0} & $\frac{3}{2}$ & $%
\frac{1}{2}$ & {\small 939} & {\small 1} & {\small 272} & $\Delta ^{+}$%
{\small (1211)} & $\Delta ^{+}${\small (1232)} & {\small 1.7\%} \\ \hline
{\small d(313)} & {\small u} & {\small d} & {\small 0} & $\frac{3}{2}$ & 
{\small -}$\frac{1}{2}$ & {\small 939} & {\small 1} & {\small 272} & $\Delta
^{0}${\small (1211)} & $\Delta ^{0}${\small (1232)} & {\small 1.7\%} \\ 
\hline
{\small d(313)} & {\small d} & {\small d} & {\small 0} & $\frac{3}{2}$ & 
{\small -}$\frac{3}{2}$ & {\small 939} & {\small 1} & {\small 272} & $\Delta
^{-}${\small (1211)} & $\Delta ^{-}${\small (1232)} & {\small 1.7\%} \\ 
\hline
{\small s(493)} & {\small u} & {\small u} & {\small -1} & {\small 1} & 
{\small 1} & {\small 1119} & {\small 1} & {\small 272} & $\Sigma ^{+}$%
{\small (1391)} & $\Sigma ^{+}${\small (1385)} & {\small 0.4\%} \\ \hline
{\small s(493)} & {\small u} & {\small d} & {\small -1} & {\small 1} & 
{\small 0} & {\small 1119} & {\small 1} & {\small 272} & $\Sigma ^{0}$%
{\small (1391)} & $\Sigma ^{0}${\small (1385)} & {\small 0.4\%} \\ \hline
{\small s(493)} & {\small d} & {\small d} & {\small -1} & {\small 1} & 
{\small -1} & {\small 1119} & {\small 1} & {\small 272} & $\Sigma ^{-}$%
{\small (1391)} & $\Sigma ^{-}${\small (1385)} & {\small 0.4\%} \\ \hline
{\small s(493)} & {\small s} & {\small u} & {\small -2} & $\frac{1}{2}$ & $%
\frac{1}{2}$ & {\small 1299} & ${\small 0}$ & {\small 204} & $\Xi ^{0}$%
{\small (1503)} & $\Xi ^{0}${\small (1532)} & {\small 1.9\%} \\ \hline
{\small s(493)} & {\small s} & {\small d} & {\small -2} & $\frac{1}{2}$ & 
{\small -}$\frac{1}{2}$ & {\small 1299} & ${\small 0}$ & {\small 204} & $\Xi
^{-}${\small (1503)} & $\Xi ^{-}${\small (1535)} & {\small 2.1\%} \\ \hline
{\small s(493)} & {\small s} & {\small s} & {\small -3} & {\small 0} & 
{\small 0} & {\small 1479} & {\small 0} & {\small 204} & $\Omega ^{-}($%
{\small 1683)} & $\Omega ^{-}(${\small 1672)} & {\small 0.7\%} \\ \hline
\end{tabular}%
\end{tabular}

\ \ \ \ \ \ \ \ \ \ \ \ \ \ \ \ \ \ \ \ \ \ \ \ \ \ \ \ \ \ \ \ \ \ \ \ \ \
\ \ \ \ \ \ \ \ \ \ \ \ \ \ \ \ \ \ \ \ \ \ \ \ \ \ \ \ \ \ \ \ \ \ \ \ \ \
\ \ \ \ \ \ \ \ 

\ \ \ \ \ \ \ \ \ \ \ \ \ \ \ \ \ \ \ \ \ \ \ \ \ \ \ \ \ \ \ \ \ \ \ \ \ \
\ \ \ \ \ \ \ \ \ \ \ \ \ \ \ \ \ \ \ \ \ \ \ \ \ \ \ \ \ \ \linebreak 
\textbf{3.3.} \textbf{Deducing} \textbf{the baryons of the SU(4) 20-plet
with SU(3) Octet }

\ \ \ \textbf{from the deduced quarks}

\ \ \ \ \ \ \ \ \ \ \ \ \ \ \ \ \ \ \ \ \ \ \ \ \ \ \ \ \ \ \ \ \ \ \ \ \ \
\ \ \ \ \ \ \ \ \ \ \ \ \ \ \ \ \ \ \ \ \ \ \ \ \ \ \ \ \ \ \ \ \ \ \ \ \ \
\ \ \ \ \ \ \ \ \ \ \ \ \ \ \ \ \ \ \ \ \ \ \ \qquad \qquad\ \ \ \ \ \ \ \ \
\ \ \ \ \ \ \ \ \ \ \ \ \ \ \ \ \qquad \qquad \qquad\ \ \ \ \ \ \ $\qquad
\qquad $

The addition of the c quark to the light quarks (u, d and s) extends the
flavor symmetry to SU(4) (basic quarks u, d, s and c). Fig 14.4 of \cite%
{Fig. 14.4} shows SU(4) multiplets of baryons. Fig 14.4 (a) shows the
20-plet baryon multiplet with the SU(3) Octet as its bottom level (see Table
6). All the baryons in a given SU(4) multiplet have the same spin and
parity. Thus from the spin and parity of the baryons in the bottom layer, we
can find the spin and parity of the 20-plet baryon multiplet:

\ \ \ \ \ \ \ \ \ \ \ \ \ \ \ \ \ \ \ \ \ \ \ 

\begin{tabular}{l}
\ Table 6: The baryons of 20-plet with SU(3) Octet ($\frac{1}{2}^{+}$) \\ 
\begin{tabular}{|l|}
\hline
Top level \ \ \ \ \ \ \ \ \ C = 2: $\Xi _{cc}^{++}$, $\Xi _{cc}^{+}$, $\
\Omega _{cc}^{+}$ \\ \hline
Middle level \ \ \ \ \ C = 1: $\Lambda _{c,}^{+}$ $\Sigma _{c}^{++}${\small %
, }$\Sigma _{c}^{+}$, $\Sigma _{c}^{0}${\small , 2}$\Xi _{c}^{+}${\small , 2}%
$\Xi _{c}^{0}${\small , }$\Omega _{c}^{0}$ \\ \hline
Bottom level \ \ \ \ C = 0: {\small p}$^{+}${\small , \ n}$^{0}${\small , \ }%
$\Lambda ^{0},$ $\Sigma ^{+},$ $\Sigma ^{0},$ $\Sigma ^{-},$ $\Xi ^{0}$, $\
\Xi ^{-}$ \\ \hline
\end{tabular}%
\end{tabular}

\ \ \ \ \ \ \ \ \ \ \ \ \ \ \ \ \ \ \ \ \ \ \ \ \ \ \ \ \ \ \ 

We have already shown the baryons of the bottom level (the SU(3) Octet of
the 20-plet) baryon multiplet with C = 0 and J$^{\text{P}}$ \ = $\frac{1}{2}%
^{+}$ in Table 4. Using the sum laws (\ref{SumB}) and the mass formula (\ref%
{Ebin of B}), we deduce the charmed number C, the strange number S, the
isospin I, the z-component I$_{z}$ of the isospin, the charge Q and the
masses of the baryons.\ \ We show the baryons on the top level in Table 7
and the middle level in Table 8 (the same name between deduced and
experimental baryons means the same quantum numbers):

\ \ \ \ \ \ \ \ \ \ \ \ \ \ \ \ \ \ \ \ \ \ \ \ \ \ \ \ \ \ \ \ \ \ \ \ \ \
\ \ \ \ \ \ \ \ \ \ 

\begin{tabular}{l}
\ \ Table 7: The top level baryons of the 20-plet (C = 2, J$^{\text{P}}$\ = $%
\frac{1}{2}^{+}$)\  \\ 
\begin{tabular}{|l|l|l|l|l|l|l|l|l|l|l|l|l|}
\hline
{\small q}$_{1}$ & {\small q}$_{2}$ & {\small q}$_{3}$ & {\small S} & 
{\small C} & {\small B} & {\small I} & {\small I}$_{z}$ & {\small M}$_{123}$
& $\Delta ${\small I} & $\Delta ${\small J} & {\small Baryon} & {\small Exp.}
\\ \hline
{\small u(313)} & {\small c} & {\small c} & {\small 0} & {\small 2} & 
{\small 0} & $\frac{1}{2}$ & $\frac{1}{2}$ & {\small 3639} & ${\small 0}$ & 
{\small 0} & $\Xi _{cc}^{++}${\small (3639)} & {\small ?} \\ \hline
{\small d(313)} & {\small c} & {\small c} & {\small 0} & {\small 2} & 
{\small 0} & $\frac{1}{2}$ & {\small -}$\frac{1}{2}$ & {\small 3639} & $%
{\small 0}$ & {\small 0} & $\Xi _{cc}^{+}${\small (3639)} & {\small ?} \\ 
\hline
{\small d}$_{s}${\small (493)} & {\small c} & {\small c} & {\small -1} & 
{\small 2} & {\small 0} & {\small 0} & {\small 0} & {\small 3819} & {\small 0%
} & {\small 0} & $\Omega _{cc}^{+}${\small (3819)} & {\small ?} \\ \hline
\end{tabular}%
\end{tabular}

\ \ \ \ \ \ \ \ \ \ \ \ \ \ \ \ \ \ \ \ \ \ \ \ \ \ \ \ \ \ \ \ \ \ \ \ \ \
\ \ \ \ \ \ \ \ \ \ \ \ \ \ \ \ \ \ \ \ 

$%
\begin{tabular}{l}
\ \ Table 8: The middle level baryons of the 20-plet{\small \ (C=1, J}$^{%
\text{P}}${\small \ = }$\frac{1}{2}^{+}${\small , }$\Delta ${\small J =0)}
\\ 
\begin{tabular}{|l|l|l|l|l|l|l|l|l|l|l|l|}
\hline
{\small q}$_{1}$ & {\small q}$_{2}$ & {\small q}$_{3}$ & {\small S} & 
{\small I} & {\small I}$_{z}$ & {\small M}$_{123}$ & $\Delta ${\small I} & $%
\Delta ${\small e} & {\small Deduced} & {\small Exper.} & $\frac{\Delta 
\text{M}}{\text{M}}{\small \%}$ \\ \hline
{\small c(1663)} & {\small u} & {\small d} & {\small 0} & {\small 0} & 
{\small 0} & {\small 2289} & {\small 0} & {\small 0} & $\Lambda _{c}^{+}$%
{\small (2289)} & $\Lambda _{c}^{+}${\small (2285)} & {\small 0.2\%} \\ 
\hline
{\small c(1663)} & {\small u} & {\small u} & {\small 0} & {\small 1} & 
{\small 1} & {\small 2289} & {\small 1} & {\small 204} & $\Sigma _{c}^{++}$%
{\small (2493)} & $\Sigma _{c}^{++}${\small (2455)} & {\small 1.5\%} \\ 
\hline
{\small c(1663)} & {\small u} & {\small d} & {\small 0} & {\small 1} & 
{\small 0} & {\small 2289} & {\small 1} & {\small 204} & $\Sigma _{c}^{+}$%
{\small (2493)} & $\Sigma _{c}^{+}${\small (2455)} & {\small 1.5\%} \\ \hline
{\small c(1663)} & {\small d} & {\small d} & {\small 0} & {\small 1} & 
{\small -1} & {\small 2289} & {\small 1} & {\small 204} & $\Sigma _{c}^{0}$%
{\small (2493)} & $\Sigma _{c}^{0}${\small (2455)} & {\small 1.5\%} \\ \hline
{\small c(1663)} & {\small u} & {\small s} & {\small -1} & $\frac{1}{2}$ & $%
\frac{1}{2}$ & {\small 2469} & {\small 0} & {\small 68} & $\Xi _{c}^{+}$%
{\small (2537)} & $\Xi _{c}^{+}${\small (}$\overline{\text{{\small 2520}}}$%
{\small )}$^{\#}$ & {\small 0.7\%} \\ \hline
{\small c(1663)} & {\small d} & {\small s} & {\small -1} & $\frac{1}{2}$ & 
{\small -}$\frac{1}{2}$ & {\small 2469} & {\small 0} & {\small 68} & $\Xi
_{c}^{0}${\small (2537)} & $\Xi _{c}^{0}${\small (}$\overline{\text{{\small %
2526}}}${\small )}$^{\#}$ & {\small 0.5\%} \\ \hline
{\small c(1663)} & {\small s} & {\small s} & {\small -2} & {\small 0} & 
{\small 0} & {\small 2649} & {\small 0} & {\small 0} & $\Omega _{c}^{0}$%
{\small (2649)} & $\Omega _{c}^{0}${\small (2698)} & {\small 1.8\%} \\ \hline
\end{tabular}
\\ 
$^{\#}\ \overline{\text{{\small 2520}}}${\small \ of }$\Xi _{c}^{+}${\small (%
}$\overline{\text{{\small 2520}}}${\small ) is the average of the mass of }$%
\Xi _{c}^{+}${\small (2466) and the mass of }$\Xi _{c}^{\prime +}${\small %
(2574)} \\ 
$^{\#}\ \overline{\text{{\small 2526}}}${\small \ of }$\Xi _{c}^{0}${\small (%
}$\overline{\text{{\small 2526}}}${\small ) is the average of the mass of }$%
\Xi _{c}^{+}${\small (2472) and the mass of }$\Xi _{c}^{\prime +}${\small %
(2579)}%
\end{tabular}%
$

\ \ \ \ \ \ \ \ \ \ \ \ \ \ \ \ \ \ \ \ \ \ \ \ \ \ \ \ \ \ \ \ \ \ \ \ \ \
\ \ \ \ \ \ \ \ \ \ \ \ \ \ \ \ \ \ \ \ \ \ \ \ \ \ \ \ \ \ \ \ \ \ \ \ \ \
\ \ \ \ \ \ \ \ \ \ \ \ \ \ \ \ \ \ 

\ \ \ \ \ \ \ \ \ \ \ \ \ \ \ \ \ \ \ \ \ \ \ \ \ \ \ \ \ \ \ \ \ \ \ \ \ \
\ \ \ \ \ \ \ \ \ \ \ \ \ \ \ \ \ \ \ \ \ \ \ \ \ \ \ \ \ \ \ \ \ \ \ \ \ \
\ \ \ \ \ \ \ \ \ \ \ \ \ \ \ \ \ \ \ \ \ \ \ \ \ \ \ \ \ \ \ \ \ \ \ \
\linebreak \textbf{3.4.}\ \textbf{Deducing} \textbf{the baryons of SU(4)
20-plet with SU(3) Decuplet }

\ \ \textbf{from the deduced quarks}

\ \ \ \ \ \ \ \ \ \ \ \ \ \ \ \ \ \ \ \ \ \ \ \ \ \ \ \ \ \ \ \ \ \ \ \ \ \
\ \ \ \ \ \ \ \ \ \ \ \ \ \ \ \ \ \ \ \qquad \qquad\ \ \ \ \ \ \ \ \ \ \ \ \
\ \ \ \ \ \ \ \ \ \ \ \ \ \ \ \ \ \ \ \ \ \ \ \ \ \ \ \ \ \ \ \ \ \ \ \ \ \
\ \ \ \ \ \ \ \ \ \ \ \ \ \ \ \ \ \ \ \ \ \ \ \ \ \ \ \ \ \ \ \textbf{\ }\ \
\ \ \ \ \ \ \ \ \ \ \qquad \qquad \qquad\ \ \ \ \ \ \ $\qquad \qquad $

Fig 14.4(b) of \cite{Fig. 14.4} shows SU(4) the 20-plet baryon multiplet
with the SU(3) Decuplet as its bottom level (see Table 9):\ 

\ \ \ \ \ \ \ \ \ \ \ \ \ \ \ \ \ \ \ \ \ \ \ \ \ \ \ \ \ \ \ \ \ \ \ \ \ \
\ \ \ 

\begin{tabular}{l}
\ \ \ \ \ Table 9: The baryons of 20-plet with Decuplet J$^{\text{P}}$ = $%
\frac{3}{2}^{+}$ \\ 
\begin{tabular}{|l|l|}
\hline
Top level \ \ \ \ \ \ \ C = 3 & \ $\Omega _{ccc}^{++}$ \\ \hline
Middle-Up\ \ \ \ \ \ C = 2 & $\Xi _{cc}^{++}$, $\Xi _{cc}^{+}$; $\ \Omega
_{cc}^{+}$ \\ \hline
Middle-Down\ \ C = 1 & $\Sigma _{c}^{++}${\small , }$\Sigma _{c}^{+}${\small %
, }$\Sigma _{c}^{0}${\small ; }$\Xi _{c}^{+}${\small , }$\Xi _{c}^{0}$%
{\small ; \ }$\Omega _{c}^{0}$ \\ \hline
Bottom \ \ \ \ \ \ \ \ \ \ C = 0 & $\Delta ^{++}$, $\Delta ^{+}$, $\Delta
^{0}$, $\Delta ^{-}$; $\Sigma ^{+}${\small , }$\Sigma ^{0}${\small , }$%
\Sigma ^{-}$: $\Xi ^{0}${\small , }$\Xi ^{-}${\small ; }$\Omega ^{-}$ \\ 
\hline
\end{tabular}%
\end{tabular}

\ \ \ \ \ \ \ \ \ \ \ \ \ \ \ \ \ \ \ \ \ \ \ \ \ \ \ \ \ \ \ \ \ \ \ \ \ \
\ \ \ \ \ \ \ \ \ \ \ 

We have already shown the baryons in the bottom level (SU(3) Decuplet) in
Table 5.\ Using sum laws (\ref{SumB}) and mass formula (\ref{Ebin of B}), we
deduce S, C, I, I$_{z}$, Q and the mass of the baryons from the S, C, I, I$%
_{z}$, Q and masses of the deduced quarks in Table 2.\ The baryons on the
middle-down level with C = 1 are shown in Table 10 (the same name between
deduced and experimental baryons means the same quantum numbers):\ 

\ \ \ \ \ \ \ \ \ \ \ \ \ \ \ \ \ \ \ \ \ \ \ \ \ \ \ \ \ \ \ \ \ \ \ \ \ \
\ \ 

$%
\begin{tabular}{l}
\ \ \ \ {\small Table 10:\ The baryons on the middle-down level with\ C = 1
and J}$^{\text{P}}${\small \ = }$\frac{3}{2}^{+}$($\Delta $J=1) \\ 
\begin{tabular}{|l|l|l|l|l|l|l|l|l|l|l|l|}
\hline
{\small q}$_{1}$ & {\small q}$_{2}$ & {\small q}$_{3}$ & {\small S} & 
{\small I} & {\small I}$_{z}$ & {\small M}$_{123}$ & $\Delta ${\small I} & $%
\Delta ${\small e} & {\small Deduced} & {\small Experim.} & $\frac{\Delta 
\text{M}}{\text{M}}${\small \%} \\ \hline
{\small c(1663)} & {\small u} & {\small u} & {\small 0} & {\small 0} & 
{\small 0} & {\small 2289} & {\small 1} & {\small 272} & $\Sigma _{c}^{++}$%
{\small (2561)} & $\Sigma _{c}^{++}${\small (2519)} & {\small 1.7\%} \\ 
\hline
{\small c(1663)} & {\small u} & {\small d} & {\small 0} & {\small 1} & 
{\small 1} & {\small 2269} & {\small 1} & {\small 272} & $\Sigma _{c}^{+}$%
{\small (2561)} & $\Sigma _{c}^{+}${\small (2515)} & {\small 1.8\%} \\ \hline
{\small c(1663)} & {\small d} & {\small d} & {\small 0} & {\small 1} & 
{\small 0} & {\small 2289} & {\small 1} & {\small 272} & $\Sigma _{c}^{0}$%
{\small (2561)} & $\Sigma _{c}^{0}${\small (2518)} & {\small 1.7\%} \\ \hline
{\small c(1663)} & {\small u} & {\small s} & {\small -1} & $\frac{1}{2}$ & $%
\frac{1}{2}$ & {\small 2469} & {\small 0} & {\small 204} & $\Xi _{c}^{+}$%
{\small (2673)} & $\Xi _{c}^{+}${\small (2645)} & {\small 1.1\%} \\ \hline
{\small c(1663)} & {\small d} & {\small s} & {\small -1} & $\frac{1}{2}$ & 
{\small -}$\frac{1}{2}$ & {\small 2469} & {\small 0} & {\small 204} & $\Xi
_{c}^{0}${\small (2673)} & $\Xi _{c}^{0}${\small (2645)} & {\small 1.1\%} \\ 
\hline
{\small c(1663)} & {\small s} & {\small s} & {\small -2} & {\small 0} & 
{\small 0} & {\small 2649} & {\small 0} & {\small 204} & $\Omega _{c}^{0}$%
{\small ( 2853 )} & $\Omega _{c}^{0}${\small ( ? )} & {\small ?} \\ \hline
\end{tabular}%
\end{tabular}%
$

\ \ \ \ \ \ \ \ \ \ \ \ \ \ \ \ \ \ \ \ \ 

The baryons on the middle-up level with C = 2 and the top level with C = 3
are shown in Table 11:

\ \ \ \ \ \ \ \ \ \ \ \ \ \ \ \ \ \ \ \ \ \ \ \ \ \ \ \ \ \ \ \ \ \ \ \ \ \
\ \ \ \ \ \ \ \ \ \ \ \ \ \ \ \ \ \ \ \ \ \ \ \ \ \ \ \ \ \ \ \ \ \ \ \ \ 

\ \ \ \ 
\begin{tabular}{l}
\ \ \ \ \ Table 11. The baryons on the middle-up and top level, J$^{\text{P}%
} $ = $\frac{3}{2}^{+}$ \\ 
\ 
\begin{tabular}{|l|l|l|l|l|l|l|l|l|l|l|l|}
\hline
{\small q}$_{1}$ & {\small q}$_{2}$ & {\small q}$_{3}$ & {\small S} & 
{\small I} & {\small I}$_{z}$ & {\small M}$_{123}$ & ${\small \Delta I}$ & $%
{\small \Delta J}$ & ${\small \Delta e}$ & {\small Deduced} & Exper. \\ 
\hline
{\small u(313)} & {\small c} & {\small c} & {\small 0} & $\frac{1}{2}$ & $%
\frac{1}{2}$ & {\small 3639} & {\small 0} & {\small 1} & {\small 204} & $%
{\small \Xi }_{cc}^{++}${\small (3843)} & ? \\ \hline
{\small d(313)} & {\small c} & {\small c} & {\small 0} & $\frac{1}{2}$ & 
{\small -}$\frac{1}{2}$ & {\small 3639} & {\small 0} & {\small 1} & {\small %
204} & ${\small \Xi }_{cc}^{+}${\small (3843)} & ? \\ \hline
{\small s(493)} & {\small c} & {\small c} & {\small 0} & {\small 0} & 
{\small 0} & {\small 3819} & {\small 0} & {\small 1} & {\small 204} & $%
{\small \Omega }_{cc}^{+}${\small (4023)} & ? \\ \hline
{\small c(1663)} & {\small c} & {\small c} & {\small 0} & {\small 0} & 
{\small 0} & {\small 4989} & {\small 0} & {\small 1} & {\small 204} & $%
{\small \Omega }_{ccc}^{++}${\small (5193)} & ? \\ \hline
\end{tabular}%
\end{tabular}%
\ \ 

\ \ \ \ \ \ \ \ \ \ \ \ \ \ \ \ \ \ \ \ \ \ \ \ \ \ \ \ \ \ \ \ \ \ \ \ \ \
\ \ \ \ \ \ \ \ \ \ \ \ \ \ \ \ \ \ \ \ \ \ \ \ \ \ \ \ \ \ \ \ \ \ \ \ \ \
\ \ \ \ \ \ \ \ \ \ \ \ \ \ \ \ \ \ \ \ \ 

Using sum laws and a phenomenological binding energy formula, in terms of q$%
\overline{\text{q}}$ meson model of the Quark Model, from the deduced quarks
in Table 2, we can deduce the masses and the quantum numbers of the
important mesons also.{\small \ }

\ \ \ \ \ \ \ \ \ \ \ \ \ \ \ \ \ \ \ \ \ \ \ \ \ \ \ \ \ \ \ \ \ \ \ \ \ \
\ \ \ \ \ \ \ \ \ \ \ \ \ \ \ 

\ \ \ \ \ \ \ \ \ \ \ \ \ \ \ \ \ \ \ \ \ \ \ \ \ \ \ \ \ \ \ \ \ \ \ \ \ \
\ \ \ \ \ \ \ \ \ \ \ \ \ \ \ \linebreak \textbf{4. \ \ Deducing the
Important Mesons from the Deduced Quarks }

\ \ \textbf{Using the q}$\overline{\text{q}}$\textbf{\ Meson Model}

\ \ \ \ \ \ \ \ \ \ \ \ \ \ \ \ \ \ \ \ \ \ \ \ \ \ \ \ \ \ \ \ \ \ \ \ \ \
\ \ \ \ \ \ \ \ \ \ \ \ \ \ \ \ \ \ \ \ \ \ \ \textbf{\ }\ \ \ \ \ \ \ \ \ \
\ \ \ \ \ \ \ \ \ \ \ \ \ \ \ \ \ \ \ \ \ \ \ \ \ \ \ \ \ \ \ \ \ \ \ \ \ \
\ \ \ \ \ \ \ \ \ \ \ \ \ \ \ \ \ \ \ \ \ \ \ \ \ \ \ \ \ \ \ \ \ \ \ \ \ \
\ \ \ \ \ \ \ \ \ \ \ \ \ \ \ \ \ \ \ \ \ \ \ \ \ \ \ \ \ \ \ \ \ \ \ \ \ \
\ \ \ \ \ \ \ \ \ \ \ \ \ \ \ \ \ \ \ \ \ \ \ \ \ \ \ \ \ \ \ \ \ \ \ \ \ \
\ \ \ \ \ \ \ \ \ \ \ \ \ \ \ \ \ \ \ \ \ \ \ \ \ \ \ \ \ \ \ \ \ \ \ \ \ \
\ \ \ \ \ \ \qquad\ \ \ \ \ \ \ \ \ \ \ \ \ \ \ \ \ \ \ \ \ \ \ \ \ \ \ \ \
\ \ \ \ \ \ \ \ \ \ \ \ \ \ \ \ \ \ \ \ \ \ \ \ \ \ \ \ \ \ \ \ \ \ \ \ \ \
\ \ \ \ \ \ \ \ \ \ \ \ \ \ \ \ \ \ \ \ \ \ \ \ \ \ \ \ \ \ \ \ \ \ \ \ \ \
\ \ \ \ \ \ \ \ \ \ \ \ \ \ \ \ \ \ \ \ \ \ \ \ \ \ \ \ \ \ \ \ \ \ \ \ \ \
\ \ \ \ \ \ \ \ \ \ \ \ \ \ \ \ \ \ \ \ \ \ \ \ \ \ \ \ \ \ \ \ \ \ \ \ \ \
\ \ \ \ \ \ \ \ \ \ \ \ \ \ \ \ \ \ \ \ \ \ \ \ \ \ \ \ \ \ \ \ \ \ \ \ \ \
\ \ \ \ \ \ \ \ \ \ \ \ \ \ \ \ \ \ \ \ \ \ \ \ \ \ \ \ \ \ \ \ \ \ \ \ \ \
\ \ \ \ \ \ \ \ \ \ \ \ \ \ \ \ \ \ \ \ \ \ \ \ \ \ \ \ \ \ \ \ \ \ \ \ \ \
\ \ \ \ \ \ \ \ \ \ \ \ \qquad\ \ \ \ \ \ \ \ \ \ \ \ \ \ \ \ \ \ \ \ \ \ \
\ \ \ \ \ \ \ \ \ \ \ \ \ \ \ \ \ \ \ \ \ \ \ \ \ \ \ \ \ \ \ \ \ \ \ \ \ \
\ \ \ \ \ \ \ \ \ \ \ \ \ \ \ \ \ \ \ \ \ \ \ \ \ \ \ \ \ \ \ \ \ \ \ \ \ \
\ \ \ \ \ \ \ \ \ \ \ \ \ \ \ \ \ \ \ \ \ \ \ \ \ \ \ \ \ \ \ \ \ \ \ \ \ \
\ \ \ \ \ \ \ \ \ \ \ \ \ \ \ \ \ \ \ \ \ \ \ \ \ \ \ \ \ \ \ \ \ \ \ \ \ \
\ \ \ \ \ \ \ \ \ \ \ \ \ \ \ \ \ \ \ \ \ \ \ \ \ \ \ \ \ \ \ \ \ \ \ \ \ \
\ \ \ \ \ \ \ \ \ \ \ \ \ \ \ \ \ \ \ \ \ \ \ \ \ \ \ \ \ \ \ \ \ \ \ \ \ \
\ \ \ \ \ \ \ \ \ \ \ \ \ \ \ \ \ \ \ \ \ \ \ \ \ \ \ \ \ \ \ \ \ \ \ \ \ \
\ \ \ \ \ \ \ \ \ \ \ \ \ \ \ \ \ \ \ \ \ \ \ \ \ \ \ \ \ \ \ \ \ \ \ \ \ \
\ \ \ \ \ \ \ \ \ \ \ \ \ \ \ \ \ \ \ \ \ \ \ \ \ \ \qquad\ \ \ \ \ \ \ \ \
\ \ \ \ \ \ \ \ \ \ \ \ \ \ \ \ \ \ \ \ \ \ \ \ \ \ \ \ \ \ \ \ \ \ \ \ \ \
\ \ \ \ \ \ \ \ \ \ \ \ \ \ \ \ \ \ \ \qquad\ \ \ \ \ \ \ \ \ \ \ \ \ \ \ \
\ \ \ \ \ \ \ \ \qquad \qquad \qquad\ \ \ \ \ \ \ \ \ \ \ \ \ \ \ \ \ \ \ \
\ \ \ \ \ \ \ \ \ \ \ \ \ \ \ \ \ \ \ \ \ \ \ \ \ \ \ \ \ \ \ \ \ \ \ \ \ \
\ \ \ \ \ \ \ \ \ \ \ \ \ \ \ \ \ \ \ \ \ \ \ \ \ \ \ \ \ \ \ \ \ \ \ \ \ \
\ \ \ \ \ \ \ \ \ \ \ \ \ \ \ \ \ \ \ \ \ \ \ \ \ \ \ \ \ \ \ \ \ \ \ \ \ \
\ \ \ \ \ \ \ \ \ \ \ \ \ \ \ \ \ \ \ \ \ \ \ \ \ \ \ \ \ \ \ \ \ \ \ \ \ \
\ \ \ \ \ \ \ \ \ \ \ \ \ \ \ \ \ \ \ \ \ \ \ \ \ \ \ \ \ \ \ \ \ \ \qquad\
\ \ \ \ \ \ \ \ \ \ \ \ \ \ \ \ \ \ \ \ \ \ \ \ \ \ \ \ \ \ \ \ \ \ \ \ \ \
\ \ \ \ \ \ \ \ \ \ \ \ \ \ \ \ \ \ \ \ \ \ \ \ \ \ \ \ \ \ \ \ \ \ \ \ \ \
\ \ \ \ \ \ \ \ \ \ \ \ \ \ \ \ \ \ \ \ \ \ \ \ \ \ \ \ \ \ \ \ \ \ \ \ \ \
\ \ \ \ \ \ \ \ \ \ \ \ \ \ \ \ \ \ \ \ \ \ \ \ \ \ \ \ \ \ \ \ \ \ \ \ \ \
\ \ \ \ \ \ \ \ \ \ \ \ \ \ \ \ \ \ \ \ \ \ \ \ \ \ \ \ \ \ \ \ \ \ \ \ \ \
\ \ \ \ \ \ \ \ \ \ \ \ \ \ \ \ \ \ \ \ \ \ \ \ \ \ \ \ \ \ \ \ \ \ \ \ \ \
\ \ \ \ \ \ \ \ \ \ \ \ \ \ \ \ \ \ \ \ \ \ \ \ \ \ \ \ \ \ \ \ \ \ \ \ \ \
\ \ \ \ \ \ \ \ \ \ \ \ \ \ \ \ \ \ \ \ \ \ \ \ \ \ \ \ \ \ \ \ \ \ \ \ \ \
\ \ \qquad\ \ \ \ \ \ \ \ \ \ \ \ \ \ \ \ \ \ \ \ \ \ \ \ \ \ \ \ \ \ \ \ \
\ \ \ \ \ \ \ \ \ \ \ \ \ \ \ \ \ \ \ \ \ \ \ \ \ \ \ \ \ \ \ \ \ \ \ \ \ \
\ \ \ \ \ \ \ \ \ \ \ \ \ \ \ \ \ \ \ \ \ \ \ \ \ \ \ \ \ \ \ \ \ \ \ \ \ \
\ \ \ \ \ \ \ \ \ \ \ \ \ \ \ \ \ \ \ \ \ \ \ \ \ \ \ \ \ \ \ \ \ \ \ \ \ \
\ \ \ \ \ \ \ \ \ \ \ \ \ \ \ \ \ \ \ \ \ \ \ \ \ \ \ \ \ \ \ \ \ \ \ \ \ \
\ \ \ \ \ \ \ \ \ \ \ \ \ \ \ \ \ \ \ \ \ \ \ \ \ \ \ \ \ \ \ \ \ \ \ \ \ \
\ \ \ \ \ \ \ \ \ \ \ \ \ \ \ \ \ \ \ \ \ \ \ \ \ \ \ \ \ \ \ \ \ \ \ \ \ \
\ \ \ \ \ \ \ \ \ \ \ \ \ \ \ \ \ \ \ \ \ \ \ \ \ \ \ \ \ \ \ \ \ \ \ \ \ \
\ \ \ \ \ \ \ \ \ \ \ \ \ \ \ \ \ \ \ \ \ \ \ \ \ \ \ \ \ \ \ \ \ \ \ \ \ \
\ \ \ \ \ \ \ \ \ \ \ \ \ \ \ \ \qquad\ \ \ \ \ \ \ \ \ \ \ \ \ \ \ \ \ \ \
\ \ \ \ \ \ \ \ \ \ \ \ \ \ \ \ \ \ \ \ \ \ \ \ \ \ \ \ \ \ \ \ \ \ \ \ \ \
\ \ \ \ \ \ \ \ \ \qquad

In the quark model, a meson is the q$_{i}\overline{\text{q}_{j}}$ bound
state of a quark q$_{i}$ and an antiquark $\overline{q_{j}}$. In this short
paper, we only deduce the important mesons to show the main physical idea
and to check the deduced quark masses. The sum laws of the baryon number $%
\mathbb{B}$, the strange number S, the charmed number C, the bottom number
B, the electric charge Q and the isospin z-component I$_{z}$ of a meson are:

\begin{equation}
\begin{tabular}{ll}
$\mathbb{B}_{\text{Meson }}$ = $\mathbb{B}_{\text{q}_{i}\text{ }}$+ $\mathbb{%
B}_{\overline{\text{q}_{j}}\text{ }}$, & S$_{\text{Meson }}$ = S$_{\text{q}%
_{i}\text{ }}$+ S$_{\overline{\text{q}_{j}}\text{ }}$, \\ 
C$_{\text{Meson }}$ = C$_{\text{q}_{i}\text{ }}$+ C$_{\overline{\text{q}_{j}}%
\text{ }}$, & B$_{\text{Meson }}$ = B$_{\text{q}_{i}\text{ }}$+ B$_{%
\overline{\text{q}_{j}}\text{ }}$, \\ 
Q$_{\text{Meson }}$ = Q$_{\text{q}_{i}\text{ }}$+ Q$_{\overline{\text{q}_{j}}%
\text{ }}$, & I$_{z,\text{Meson }}$ = I$_{z,\text{q}_{i}\text{ }}$+I$_{z,%
\text{q}_{j}\text{ }}$.%
\end{tabular}
\label{Sum of Meson}
\end{equation}%
The masses of the quarks are huge from the term $\Delta $ ($\gg $ M$_{\text{p%
}}$ = 938 Mev) of the quark's masses in Table 2. The masses of mesons \cite%
{Meson}, however, are not huge. Thus we infer that there will be (-2$\Delta $%
) in the binding energy of the meson to cancel 2$\Delta $ in the masses of
the quark and antiquark inside the meson. We assume a phenomenological
binding energy formula 
\begin{equation}
\text{E}_{M}\text{(q}_{i}\overline{q_{j}}\text{) = - 2}\Delta \text{ - 338
+100[}\frac{\Delta m}{\text{{\small 938}}}\text{ +C}_{ij}\text{ -}\left\vert 
\text{SC}\right\vert \text{ -2.5}\left\vert \text{B}\right\vert \text{- }%
\delta (\Delta \text{m})\text{\ -2I}_{i}\text{I}_{j}\text{] ,}
\label{Bind EM}
\end{equation}%
where $\Delta $m =$\left\vert \text{m}_{q_{i}}\text{-m}_{\overline{q_{j}}%
}\right\vert $; C$_{ij}$ = C$_{_{i}}$ - $\overline{\text{C}_{j}}$; $%
\left\vert \text{SC}\right\vert $ is the absolute value of strange number $%
\times $ charmed number of the meson; $\left\vert \text{B}\right\vert $ is
the absolute value of the bottom number of the meson.$\ \ \delta (\Delta $m)
is Dirac $\delta $ function; for $\Delta $m = 0, $\delta (\Delta $m) = 1;
for $\Delta $m $\neq $ 0, $\delta (\Delta $m) = 0. I$_{i}$ is the isospin of
q$_{i}$; I$_{j}$ is the isospin of q$_{j}$.\ The meson mass formula is%
\begin{equation}
\text{M}_{\text{Meson}}\text{ = m}_{\text{q}_{i}}\text{ + m}_{\overline{%
\text{q}_{j}}}\text{ + E}_{M}\text{(q}_{i}\overline{\text{q}_{j}}\text{).}
\label{Meson Mass}
\end{equation}

Using (\ref{Meson Mass}), (\ref{Bind EM}) and sum laws (\ref{Sum of Meson}),
from the deduced quark masses in Table 2, we can deduce the masses of the
important mesons as shown in the following. Since (-2$\Delta $) in the
binding energy of the meson is always cancelled by 2$\Delta $ in the masses
of the quark and antiquark inside the meson, we can omit the (-2$\Delta $)
inside (\ref{Bind EM}) and the 2$\Delta $ inside the masses of the quark and
antiquark when we deduce masses of mesons.

\ \ \ \ \ \ \ \ \ \ \ \ \ \ \ \ \ \ \ \ \ \ \ \ \ \ \ \ \ \ \ \ \ \ \ \ \ \
\ \ \ \ 

\ \ \ \ \ \ \ \ \ \ \ \ \ \ \ \ \ \ \ \ \ \ \ \ \ \ \ \ \ \ \ \ \ \ \ \ \ \
\ \ \ \ \ \ \ \ \ \ \ \ \ \ \ \ \ \ \ \ \ \ \ \ \ \ \ \ \ \ \ \ \ \ \ \ \ \
\ \ \ \ \ \ \ \ \ \ \ \ \ \ \ \ \ \ \ \ \ \ \ \ \ \ \ \ \ \ \ \ \ \ \ \ \ \
\ \ \ \ \ \ \ \ \ \ \ \ \ \ \ \ \ \ \ \ \ \ \ \ \ \ \ \ \ \ \ \ \ \linebreak 
\textbf{4.1. Deducing the mesons\ made of quarks and their own antiquarks }

\ \ \ \textbf{from the deduced quarks}

\ \ \ \ \ \ \ \ \ \ \ \ \ \ \ \ \ \ \ \ \ \ \ \ \ \ \ \ \ \ \ \ \ \ \ \ \ \
\ \ \ \ \ \ \ \ \ \ \ \ \ \ \ \ \ \ \ \ \ \ \ \ \ \ \ \ \ \ \ \ \ \ 

For these mesons (i = j), $\Delta $m = 0, $\delta (\Delta m)$ = 1, $%
\left\vert \text{SC}\right\vert $ = 0, $\left\vert \text{B}\right\vert $ =
0, the binding energy formula (\ref{Bind EM}) is simplified into\ \ \ 
\begin{equation}
\text{E}_{M}\text{(q}_{i}\overline{q_{i}}\text{) = - 438 +100(C}_{ii}\text{\
- 2I}_{i}\text{I}_{i}\text{).}  \label{Pair}
\end{equation}%
Using sum laws (\ref{Sum of Meson}), the mass formula (\ref{Meson Mass}) and
the binding energy formula (\ref{Pair}), we deduce the mesons made of the
quarks and their own antiquarks as shown in Table 12 (the same names show
the same quantum numbers of the deduced and experimental mesons):

\ \ \ \ \ \ \ \ \ \ \ \ \ \ \ \ \ \ \ \ \ \ \ \ \ \ \ \ \ \ \ \ \ \ \ \ \ \
\ \ \ \ \ 

\begin{tabular}{l}
$\ \ \text{Table 12.\ \ The Most Important Mesons ({\small B}}${\small aryon
number\ }$\mathbb{B}${\small \ = 0}) \\ 
$%
\begin{tabular}{|l|l|l|l|l|l|}
\hline
$\text{q}_{i}\text{(m}_{i}\text{)}\ \overline{\text{q}_{j}\text{(m}_{j}\text{%
)}}$ & $\text{C}_{ij}$ & E$_{M}$(q$_{i}\overline{\text{{\small q}}_{j}}$) & 
Deduced & Exper. & R \\ \hline
$\text{q}_{N}$(313)$\overline{\text{q}_{N}\text{(313)}}$ & 0 & - 488 & $\pi $%
(138)$^{\#}$ & $\pi $($\overline{\text{137}}$)$^{\#}$ & 0.7 \\ \hline
s(493)$\overline{\text{s(493)}}$ & 0 & - 438 & $\eta $(548) & $\eta $(548) & 
0.0 \\ \hline
c(1663)$\overline{\text{c(1663)}}$ & 2 & - 238 & J/$\psi $(3088) & J/$\psi $%
(3097) & 0.3 \\ \hline
b(4993)$\overline{\text{b(4993)}}$ & 0 & -\ 438 & $\Upsilon $(9548) & $%
\Upsilon $(9460) & 0.9 \\ \hline
c$^{\ast }$(5893)$\overline{\text{c}^{\ast }\text{(5893)}}$ & 2 & - 238 & $%
\psi ^{\ast }$(11548)$^{\$}$ & \ \ \ \ \ ? &  \\ \hline
c$^{\ast }$(13003)$\overline{\text{c}^{\ast }\text{(13003)}}$ & 2 & - 238 & $%
\psi ^{\ast }$(25768)$^{\$}$ & \ \ \ \ \ ? &  \\ \hline
b$^{\ast }$(15253)$\overline{\text{b}^{\ast }\text{(15253)}}$ & 0 & - 438 & $%
\Upsilon ^{\ast }$(29135)$^{\$}$ & \ \ \ \ \ ? &  \\ \hline
\end{tabular}%
$ \\ 
$^{\$}$ $\psi ^{\ast }${\small (11548)}$^{\$}${\small , }$\psi ^{\ast }$%
{\small (25768)}$^{\$}${\small \ and }$\Upsilon ^{\ast }${\small (29135)}$%
^{\$}${\small \ might be only quasi-mesons}%
\end{tabular}

\ \ \ \ \ \ \ \ \ \ \ \ \ \ \ \ \ \ \ \ \ 

In Table 12, the first row $\text{q}_{N}$(313)$\overline{\text{q}_{N}\text{%
(313)}}$ = $\pi $(138)$^{\#}$ is deduced using Table 13. The quark $\text{q}%
_{N}$(313) represents the quarks u(313) and d(313) with I = $\frac{1}{2}$%
(see Table2). The mass $\overline{\text{137}}$ of $\pi $($\overline{\text{137%
}}$) is the average mass of $\pi ^{+}$(140), $\pi ^{-}$(140) and $\pi ^{0}$%
(135):

\ \ \ \ \ \ \ \ \ \ \ \ \ \ \ \ \ \ \ \ \ \ \ \ \ \ \ \ \ \ \ \ \ \ \ \ \ \
\ \ \ 

\begin{tabular}{l}
\ \ \ \ \ \ \ \ \ \ \ \ \ \ Table 13. \ The Deduction of $\pi $(138)$^{\#}$
and $\pi $($\overline{\text{137}}$)$^{\#}$ \  \\ 
\begin{tabular}{|l|l|l|l|l|l|l|}
\hline
$\text{q}_{i}\text{(m}_{i}\text{)}\ \overline{\text{q}_{j}\text{(m}_{j}\text{%
)}}$ & $\text{C}_{ij}\text{\ }$ & -438 & 100$\times \text{2I}_{i}\text{I}_{%
\text{j}}$ & E$_{bind}$ & Deduced & Exper. \\ \hline
u(313)$\overline{\text{d(313)}}$ & 0 & -438 & \ \ \ 50 & - 488 & $\pi ^{+}$%
(138) & $\pi ^{+}$(140) \\ \hline
d(313)$\overline{\text{u(313)}}$ & 0 & -438 & \ \ \ 50 & - 488 & $\pi ^{-}$%
(138) & $\pi ^{-}$(140) \\ \hline
u(313)$\overline{\text{u(313)}}$ & 0 & -438 & \ \ \ 50 & - 488 & $\pi ^{0}$%
(138) & $\pi ^{0}$(135) \\ \hline
d(313)$\overline{\text{d(313)}}$ & 0 & -438 & \ \ \ 50 & - 488 & $\pi ^{0}$%
(138) & $\pi ^{0}$(135) \\ \hline
$\text{q}_{N}$(313)$\overline{\text{q}_{N}\text{(313)}}$ & 0 & -438 & \ \ \
50 & - 488 & $\pi $(138)$^{\#}$ & $\pi $($\overline{\text{137}}$) \\ \hline
\end{tabular}%
\end{tabular}

\ \ \ \ \ \ \ \ \ \ \ \ \ \ \ \ \ \ \ \ \ \ \ \ \ \ \ \ \ \ \ \ \ \ \ \ \ \
\ \ \ \ \ 

Table 12 shows that the deduced masses of the most important mesons are more
than 99\% consistent with the experimental results \cite{Meson} and the
deduced quantum numbers match the experimental result exactly (the same
names show the same quantum numbers of the deduced and experimental mesons).
These results might show that the deduced quark masses are really correct. $%
\psi ^{\ast }$(11548)$^{\$}$, $\psi ^{\ast }$(25768)$^{\$}$ and $\Upsilon
^{\ast }$(29135)$^{\$}$ are not normal mesons. They\ might be quasi-mesons,
and we are not sure that they really exist in nature.

\ \ \ \ \ \ \ \ \ \ \ \ \ \ \ \ \ \ \ \ \ \ \ \ \ \ \ \ \ \ \ \ \ \ \ \ \ \
\ \ \ \ \ \ \ \ \ \ \ \ \ \ \ \ \ \ \ \ \ \ \ \ \ \ \ \ \ \ \ \ \ \ \ \ \ \
\ \ \ \linebreak \textbf{4.2. Deducing the mesons\ made of quarks and other
antiquarks\ (J}$^{\text{P}}$\textbf{= 0}$^{\text{\textbf{-}}}$\textbf{) }

\ \ \ \textbf{from the deduced quarks}

\ \ \ \ \ \ \ \ \ \ \ \ \ \ \ \ \ \ \ \ \ \ \ \ \ \ \ \ \ \ \ \ \ \ \ \ \ \
\ \ \ \ \ \ \ \ \ \ \ \ \ \ \ \ \ \ \ \ \ \ \ \ \qquad \qquad\ \ \ \ \ \ \ \
\ \ \ \ \ \ \ \ \ \ \ \ \ \ \ \ \ \ \ \ \ \ \ \ \ 

For these mesons (i $\neq $ j) ; $\Delta $m $\neq $ 0 $\rightarrow \delta $($%
\Delta $m) = 0; and i $\neq $ j$\rightarrow $ 2I$_{i}$I$_{j}$ = 0 since
there is only one kind of quark q$_{N}$(313) with I = $\frac{1}{2}$, and all
other kinds of quarks have I = 0.\ The binding energy formula (\ref{Bind EM}%
) is simplified into\ \ 

\begin{equation}
\text{E}_{M}\text{(q}_{i}\overline{q_{j}}\text{) = - 338+100[}\frac{\Delta m%
}{\text{{\small 938}}}\text{+ C}_{ij}\text{ }{\small -}\left\vert \text{SC}%
\right\vert \text{ - 2.5}\left\vert \text{B}\right\vert \text{ ].}
\label{UnPair}
\end{equation}%
Using the sum laws (\ref{Sum of Meson}), the binding energy (\ref{UnPair})
and the meson mass formula (\ref{Meson Mass}), from the deduced quarks in
Table 2, we deduce the mesons shown in Table 14.

Table 14 shows that the deduced quantum numbers (I, S, C, B and Q) of the
mesons are the same as the experimental result (the same names show the same
quantum numbers of the deduced and experimental mesons),\ and the deduced
masses of the important pseudoscalar mesons are about 98\% consistent with
the experimental results. These results furthermore show that the deduced
masses and quantum numbers might be indeed correct.\ 

\ \ \ \ \ \ \ \ \ \ \ \ \ \ \ \ \ \ \ \ \ \ \ \ \ \ \ \ \ \ \ \ \ \ \ \ \ \
\ \ \ \ \textbf{\ }

$%
\begin{tabular}{l}
\ \ \ \ \ Table 14. The Pseudoscalar Mesons ($\mathbb{B}$ =0, J$^{\text{P}}$%
=0$^{-\text{ }}$, R = $\frac{\text{M}_{D}\text{-M}_{E}}{\text{ \ \ M \ \ }}$%
\%) \\ 
$%
\begin{tabular}{|l|l|l|l|l|l|l|l|l|l|l|}
\hline
{\small q}$_{i}\overline{q_{j}}$ & {\small S} & {\small C} & {\small B} & 
{\small Q} & $\frac{{\small \Delta m}}{\text{{\small 9}{\LARGE .}{\small 38}}%
}$ & {\small C}$_{ij}$ & {\small E}$_{M}$ & {\small Deduced} & {\small Exper.%
} & {\small R} \\ \hline
{\small u(313)}$\overline{\text{s{\small (493)}}}$ & {\small 1} & {\small 0}
& {\small 0} & {\small 1} & {\small 19} & {\small 0} & {\small -319} & 
{\small K}$^{+}${\small (487)} & {\small K}$^{+}${\small (494)} & {\small 1.4%
} \\ \hline
{\small d(313)}$\overline{\text{s{\small (493)}}}$ & {\small 1} & {\small 0}
& {\small 0} & {\small 0} & {\small 19} & {\small 0} & {\small -319} & 
{\small K}$^{0}${\small (487)} & {\small K}$^{0}${\small (498)} & {\small 2.2%
} \\ \hline
{\small s(493)}$\overline{\text{{\small u(313)}}}$ & {\small -1} & {\small 0}
& {\small 0} & {\small -1} & {\small 19} & {\small 0} & {\small -319} & 
{\small K}$^{-}${\small (487)} & {\small K}$^{-}${\small (494)} & {\small 1.4%
} \\ \hline
{\small s(493)}$\overline{\text{{\small d(313)}}}$ & {\small -1} & {\small 0}
& {\small 0} & {\small 0} & {\small 19} & {\small 0} & {\small -319} & $%
\overline{\text{{\small K}}^{0}}${\small (487)} & $\overline{\text{{\small K}%
}^{0}}${\small (498)} & {\small 2.2} \\ \hline
{\small c(1663)}$\overline{\text{{\small d(313)}}}$ & {\small 0} & {\small 1}
& {\small 0} & {\small 1} & {\small 144} & {\small 1} & {\small -94} & 
{\small D}$^{+}${\small (1882)} & {\small D}$^{+}${\small (1869)} & {\small %
0.7} \\ \hline
{\small c(1663)}$\overline{\text{{\small u(313)}}}$ & {\small 0} & {\small 1}
& {\small 0} & {\small 0} & {\small 144} & {\small 1} & {\small -94} & 
{\small D}$^{0}${\small (1882)} & {\small D}$^{0}${\small (1865)} & {\small %
0.9} \\ \hline
{\small c(1663)}$\overline{\text{s{\small (493)}}}$ & {\small 1} & {\small 1}
& {\small 0} & {\small 1} & {\small 125} & {\small 1} & {\small -213} & 
{\small D}$_{s}^{+}${\small (1943)} & {\small D}$_{s}^{+}${\small (1968)} & 
{\small 1.3} \\ \hline
{\small d(313)}$\overline{\text{c{\small (1663)}}}$ & {\small 0} & {\small -1%
} & {\small 0} & {\small -1} & {\small 144} & {\small 1} & {\small -94} & 
{\small D}$^{-}${\small (1882)} & {\small D}$^{-}${\small (1869)} & {\small %
0.7} \\ \hline
{\small u(313)}$\overline{\text{c{\small (1663)}}}$ & {\small 0} & {\small -1%
} & {\small 0} & {\small 0} & {\small 144} & {\small 1} & {\small -94} & $%
\overline{\text{{\small D}}^{0}}${\small (1882)} & $\overline{\text{{\small D%
}}^{0}}${\small (1865)} & {\small 0.9} \\ \hline
{\small s(493)}$\overline{\text{c{\small (1663)}}}$ & {\small -1} & {\small %
-1} & {\small 0} & {\small -1} & {\small 125} & {\small 1} & {\small -213} & 
$\overline{\text{{\small D}}_{s}}^{-}${\small (1943)} & $\overline{\text{%
{\small D}}_{s}}^{-}${\small (1968)} & {\small 1.3} \\ \hline
{\small u(313)}$\overline{\text{b{\small (4993)}}}$ & {\small 0} & {\small 0}
& {\small 1} & {\small 1} & {\small 499} & 0 & {\small -89} & {\small B}$%
^{+} ${\small (5217)} & {\small B}$^{+}${\small (5279)\ } & {\small 1.2} \\ 
\hline
{\small b(4993)}$\overline{\text{u{\small (313)}}}$ & {\small 0} & {\small 0}
& {\small -1} & {\small -1} & {\small 499} & 0 & {\small -89} & {\small B}$%
^{-}${\small (5217)\ } & {\small B}$^{-}${\small (5279)\ } & {\small 1.2} \\ 
\hline
{\small d(313)}$\overline{\text{b{\small (4993)}}}$ & {\small 0} & {\small 0}
& {\small 1} & {\small 0} & {\small 499} & 0 & {\small -89} & $\overline{%
\text{{\small B}}}^{0}${\small (5217)} & $\overline{\text{{\small B}}}^{0}$%
{\small (5279)} & {\small 1.2} \\ \hline
{\small b(4993)}$\overline{\text{d{\small (313)}}}$ & {\small 0} & {\small 0}
& {\small -1} & {\small 0} & {\small 499} & 0 & {\small -89} & {\small B}$%
^{0}${\small (5217)} & {\small B}$^{0}${\small (5279)} & {\small 1.2} \\ 
\hline
{\small b(4993)}$\overline{\text{{\small s(493)}}}$ & {\small 1} & {\small 0}
& {\small -1} & {\small 0} & {\small 480} & 0 & {\small -108} & {\small B}$%
_{s}^{0}${\small (5378)} & {\small B}$_{s}^{0}${\small (5370)} & {\small 0.2}
\\ \hline
{\small s(493)}$\overline{\text{b{\small (4993)}}}$ & {\small -1} & {\small 0%
} & {\small 1} & {\small 0} & {\small 480} & 0 & {\small -108} & $\overline{%
\text{{\small B}}}_{s}^{0}${\small (5378)} & $\overline{\text{{\small B}}}%
_{s}^{0}${\small (5370)} & {\small 0.2} \\ \hline
{\small c(1663)}$\overline{\text{b{\small (4993)}}}$ & {\small 0} & {\small 1%
} & {\small 1} & {\small 1} & {\small 355} & {\small 1} & {\small -133} & 
{\small B}$_{c}^{+}${\small (6523)} & {\small B}$_{c}^{+}${\small (6400)} & 
{\small 1.9} \\ \hline
{\small b(4993)}$\overline{\text{c{\small (1663)}}}$ & {\small 0} & {\small %
-1} & {\small -1} & {\small -1} & {\small 355} & {\small 1} & {\small -133}
& {\small B}$_{c}^{-}${\small (6523)} & {\small B}$_{c}^{-}${\small (6400)}
& {\small 1.9} \\ \hline
\end{tabular}%
$%
\end{tabular}%
$

\ \ \ \ \ \ \ \ \ \ \ \ \ \ \ \ \ \ \ \ \ \ \ \ \ \ \ \ \ \ \ \ \ \ \ \ \ \
\ \ \ \ \ \ \ \ \ \ \ \ \ \ \ \ \ \ \ \ \ \ \ \ \ \ \ \ \ \ \ \ \ \ \ \ \ \
\ \ \ \ \ \ \ \ \ \ \ \ \ \ \ \ \ \ \ \ \ \ \ \ \ \ \ \ \ \ \ \ \ \ \ \ 

\ \ \ \ \ \ \ \ \ \ \ \ \ \ \ \ \ \ \ \ \ \ \ \ \ \ \ \ \ \ \ \ \ \ \ \ \ \
\ \ \ \ \ \ \ \ \ \ \ \ \ \ \ \textbf{\ }\linebreak \textbf{5. \ Discussion\
and Predictions}

\ \ \ \ \ \ \ \ \ \ \ \ \ \ \ \ \ \ \ \ \ \ \ \ \ \ \ \ \ \ \ \ \ \ \ \ \ \
\ \ \ \ \ \ \ \ \ \ \ \ \ \ \ \ \ \ \ \ \ \ \ \ \ \ \ \ \ \ \ \ \ \ \ \ \ \
\ \ \ \ \ \ \ \ \ \ \ \ \ \ \ \ \ \ \ \ \ \ \ \ \ \ \ \ \ \ \ \ \ \ \ \ \ \
\ \ \ \ \ \ \ \ \ \ \ \ \ \ \ \ \ \ \ \ \ \ \ \ \ \ \ \ \ \ \ \ \ \ \ \ \ \
\ \ \ \ \ \ \ \ \ \ \ \ \ \ \ \ \ \ \ \ \ \ \ \ \ \ \ \ \ \ \ \ \ \ \ \ \ \
\ \ \ \ \ \ \ \ \ \ \ \ \ \ \ \ \ \ \ \ \ \ \ \ \ \ \ \ \ \ \ \ \ \ \ \ \ \
\ \ \ \ \ \ \ \ \ \ \ \ \ \ \ \ \ \ \ \ \ \ \ \ \ \ \ \ \ \ \ \ \ \ \ \ \ \
\ \ \ \ \ \ \ \ \ \ \ \ \qquad\ \ \ \ \ \ \ \ \ \ \ \ \ \ \ \ \ \ \ \ \ \ \
\ \ \ \ \ \ \ \ \ \ \ \ \ \ \ \ \ \ \ \ \ \ \ \ \ \ \ \ \ \ \ \ \ \ \ \ \ \
\ \ \ \ \ \ \ \ \ \ \ \ \ \ \ \ \ \ \ \ \ \ \ \ \ \ \ \ \ \ \ \ \ \ \ \ \ \
\ \ \ \ \ \ \ \ \ \ \ \ \ \ \ \ \ \ \ \ \ \ \ \ \ \ \ \ \ \ \ \ \ \ \ \ \ \
\ \ \ \ \ \ \ \ \ \ \ \ \ \ \ \ \ \ \ \ \ \ \ \ \ \ \ \ \ \ \ \ \ \ \ \ \ \
\ \ \ \ \ \ \ \ \ \ \ \ \ \ \ \ \ \ \ \ \ \ \ \ \ \ \ \ \ \ \ \ \ \ \ \ \ \
\ \ \ \ \ \ \ \ \ \ \ \ \ \ \ \ \ \ \ \ \ \ \ \ \ \ \ \ \ \ \ \ \ \ \ \ \ \
\ \ \ \ \ \ \ \ \ \ \ \ \ \ \ \ \ \ \ \ \ \ \ \ \ \ \ \ \ \ \ \ \ \ \ \ \ \
\ \ \ \ \ \ \ \ \ \ \ \ \ \ \ \ \ \ \ \ \ \ \ \ \ \ \ \ \ \ \ \ \ \ \ \ \ \
\ \ \ \ \ \ \ \ \ \ \ \ \ \ \ \ \ \ \ \ \ \ \ \ \ \ \qquad\ \ \ \ \ \ \ \ \
\ \ \ \ \ \ \ \ \ \ \ \ \ \ \ \ \ \ \ \ \ \ \ \ \ \ \ \ \ \ \ \ \ \ \ \ \ \
\ \ \ \ \ \ \ \ \ \ \ \ \ \ \ \ \ \ \ \ \ \ \ \ \ \ \ \qquad \qquad \qquad\
\ \ \ \ \ \ \ \ \ \ \ \ \ \ \ \ \ \ \ \ \ \ \ \ \ \ \ \ \ \ \ \ \ \ \ \ \ \
\ \ \ \ \ \ \ \ \ \ \ \ \ \ \ \ \ \ \ \ \ \ \ \ \ \ \ \ \ \ \ \ \ \ \ \ \ \
\ \ \ \ \ \ \ \ \ \ \ \ \ \ \ \ \ \ \ \ \ \ \ \ \ \ \ \ \ \ \ \ \ \ \ \ \ \
\ \ \ \ \ \ \ \ \ \ \ \ \ \ \ \ \ \ \ \ \ \ \ \ \ \ \ \ \ \ \ \ \ \ \ \ \ \
\ \ \ \ \ \ \ \ \ \ \ \ \ \ \ \ \ \ \ \ \ \ \ \ \ \ \ \ \ \ \ \ \ \ \ \ \ \
\ \ \ \ \ \ \ \ \ \ \ \ \ \ \ \qquad\ \ \ \ \ \ \ \ \ \ \ \ \ \ \ \ \ \ \ \
\ \ \ \ \ \ \ \ \ \ \ \ \ \ \ \ \ \ \ \ \ \ \ \ \ \ \ \ \ \ \ \ \ \ \ \ \ \
\ \ \ \ \ \ \ \ \ \ \ \ \ \ \ \ \ \ \ \ \ \ \ \ \ \ \ \ \ \ \ \ \ \ \ \ \ \
\ \ \ \ \ \ \ \ \ \ \ \ \ \ \ \ \ \ \ \ \ \ \ \ \ \ \ \ \ \ \ \ \ \ \ \ \ \
\ \ \ \ \ \ \ \ \ \ \ \ \ \ \ \ \ \ \ \ \ \ \ \ \ \ \ \ \ \ \ \ \ \ \ \ \ \
\ \ \ \ \ \ \ \ \ \ \ \ \ \ \ \ \ \ \ \ \ \ \ \ \ \ \ \ \ \ \ \ \ \ \ \ \ \
\ \ \ \ \ \ \ \ \ \ \ \ \ \ \ \ \ \ \ \ \ \ \ \ \ \ \ \ \ \ \ \ \ \ \ \ \ \
\ \ \ \ \ \ \ \ \ \ \ \ \ \ \ \ \ \ \ \ \ \ \ \ \ \ \ \ \ \ \ \ \ \ \ \ \ \
\ \ \ \ \ \ \ \ \ \ \ \ \ \ \ \ \ \ \ \ \ \qquad\ \ \ \ \ \ \ \ \ \ \ \ \ \
\ \ \ \ \ \ \ \ \ \ \ \ \ \ \ \ \ \ \ \ \ \ \ \ \ \ \ \ \ \ \ \ \ \ \ \ \ \
\ \ \ \ \ \ \ \ \ \ \ \ \ \ \ \ \ \ \ \ \ \ \ \ \ \ \ \ \ \ \ \ \ \ \ \ \ \
\ \ \ \ \ \ \ \ \ \ \ \ \ \ \ \ \ \ \ \ \ \ \ \ \ \ \ \ \ \ \ \ \ \ \ \ \ \
\ \ \ \ \ \ \ \ \ \ \ \ \ \ \ \ \ \ \ \ \ \ \ \ \ \ \ \ \ \ \ \ \ \ \ \ \ \
\ \ \ \ \ \ \ \ \ \ \ \ \ \ \ \ \ \ \ \ \ \ \ \ \ \ \ \ \ \ \ \ \ \ \ \ \ \
\ \ \ \ \ \ \ \ \ \ \ \ \ \ \ \ \ \ \ \ \ \ \ \ \ \ \ \ \ \ \ \ \ \ \ \ \ \
\ \ \ \ \ \ \ \ \ \ \ \ \ \ \ \ \ \ \ \ \ \ \ \ \ \ \ \ \ \ \ \ \ \ \ \ \ \
\ \ \ \ \ \ \ \ \ \ \ \ \ \ \ \ \ \ \ \ \ \ \ \ \ \ \ \ \ \ \ \ \ \ \ \ \ \
\ \ \ \ \ \ \ \ \ \ \ \ \ \ \ \ \ \ \ \ \ \ \ \ \ \ \ \ \ \ \ \ \ \ \
\qquad\ \ \ \ \ \ \ \ \ \ \ \ \ \ \ \ \ \ \ \ \ \ \ \ \ \ \ \ \ \ \ \ \ \ \
\ \ \ \ \ \ \ \ \ \ \ \ \ \ \ \ \ \ \ \ \ \ \ \ \ \ \ \ \ \ \ \qquad

1). The extended Planck-Bohr quantization (\ref{Qk-Mass}) plays a crushing
role in deducing the flavored quarks. Without this quantization, from two
elementary quarks $\epsilon _{\text{u}}$ and $\epsilon _{\text{d}}$ in the
vacuum state, we cannot deduce the flavored excited quarks; we only can
deduce the normally excited quarks u and d. The flavored quarks (s, c and b)
are the results of the extended Planck-Bohr quantization.

2). The fact that physicists have not found any free quark shows that the
binding energies of baryons and mesons are huge. The binding energy (-3$%
\Delta $) of baryons (or -2$\Delta $ of mesons) is a phenomenological
approximation of the interaction energy in a baryon (or a meson). The
binding energy (-3$\Delta $) (or -2$\Delta $) is always cancelled by the
corresponding parts (3$\Delta $) (or 2$\Delta $) of the masses of the three
quarks inside the baryon (or the quark and antiquark inside the meson). Thus
we can omit the binding energy and the corresponding mass parts of the three
quarks (or the quark and antiquark) when we account the mass of the baryon
(or the meson). This effect makes it appear as if there is no binding energy
in the baryon (or the meson). We, however, cannot always forget the huge
binding energy (-3$\Delta $ for baryon) (or -2$\Delta $ for meson). In fact
the huge binding energy is a necessary condition for the quark confinement.

3). Since the mass of the top quark is too large (about 185 times proton
mass),\ using these phenomenological formulae, we cannot deduce it. This
paper tries to deduce the masses of baryons and mesons (there are
experimental results that can compare with the deduced results). Because
there is not any baryon or meson that contains the top quark, we do not need
the mass of the top quark in this paper. How to deduce the very large mass
of the top quark is still an open problem of this paper.

4). This paper predicts some baryons shown in Table 15:

\ \ \ \ \ \ \ \ \ \ \ \ \ \ \ \ \ \ \ \ \ \ \ \ \ \ \ \ \ \ \ \ \ \ \ \ \ \
\ \ \ \ \ 

\begin{tabular}{l}
\ \ \ \ \ \ \ \ \ \ \ \ \ \ \ \ \ \ \ \ \ \ Table 15. \ The Predicted
Baryons and Mesons \\ 
\begin{tabular}{|l|l|l|l|l|l|}
\hline
Baryon J$^{\text{P}}$\ = $\frac{1}{2}^{+}$ & $\Xi _{cc}^{++}${\small (3639)}
& $\Xi _{cc}^{+}${\small (3639)} & $\Omega _{cc}^{+}${\small (3819)} &  & 
\\ \hline
Baryon J$^{\text{P}}$ = $\frac{3}{2}^{+}$ & ${\small \Xi }_{cc}^{++}${\small %
(3843)} & ${\small \Xi }_{cc}^{+}${\small (3843)} & $\Omega _{c}^{0}${\small %
(2853)} & ${\small \Omega }_{cc}^{+}${\small (4023)} & ${\small \Omega }%
_{ccc}^{++}${\small (5193)} \\ \hline
\end{tabular}%
\end{tabular}

\ \ \ \ \ \ \ \ \ \ \ \ \ \ \ \ \ \ \ \ \ \ \ \ \ \ \ \ \ \ \ \ \ \ \ \ \ \
\ \ \ \ \ \ \ \ \ \ 

\ The predicted baryons have higher energies than the discovered baryons or
higher flavor values such as C = 2 or 3; they are difficult to discover.\ As
new experimental technology and equipments become available, these predicted
baryons might be more easily discovered. At the same time, a quasi-baryon $%
\Lambda _{c^{\ast }}^{\ast }$(6519) [c$^{\ast }$(5893)u(313)d(313)] and the
quasi-mesons $\psi ^{\ast }$(11548), $\psi ^{\ast }$(25768) and $\Upsilon
^{\ast }$(29135) might have a slight possibility of being discovered using
the new experimental technology and equipments. We are not sure that they
really exist in nature.

\ \ \ \ \ \ \ \ \ \ \ \ \ \ \ \ \ \ \ \ \ \ \ \ \ \ \ \ \ \ \ \ \ \ \ \ \ \
\ \ \ \ \ \ \ \ \ \ \ \ \ \ \ \ 

\ \ \ \ \ \ \ \ \ \ \ \ \ \ \ \ \ \ \ \ \ \ \ \ \ \ \ \ \ \ \ \ \ \ \ \ \ \
\ \ \ \linebreak \textbf{6. \ Conclusions}

\ \ \ \ \ \ \ \ \ \ \ \ \ \ \ \ \ \ \ \ \ \ \ \ \ \ \ \ \ \ \ \ \ \ \ \ \ \
\ \ \ \ \ \ \ \ \ \ \ \ \ \ \ \ \ \ \ \ \ \ \ \ \ \ \ \ \ \ \ \ \ \ \ \ \ \
\ \ \ \ \ \ \ \ \ \ \ \ \ \ \ \ \ \ \ \ \ \ \ \ \ \ \ \ 

\ 1). There might be only one kind of unflavored (S = C = B = 0) elementary
quark family $\epsilon $ with three possible colors (red or blue or yellow),
the baryon number $\mathbb{B}$ = $\frac{1}{3}$, the spin s = $\frac{1}{2}$,
the isospin I = $\frac{1}{2}$ and two isospin states ($\epsilon _{u}$ with I$%
_{z}$ = $\frac{1}{2}$ and Q = +$\frac{2}{3}$, and $\epsilon _{d}$ with I$%
_{z} $ = -$\frac{1}{2}$ and Q = -$\frac{1}{3}$). Thus there are six
elementary quarks in the elementary quark family.

2). The elementary quarks are essentially in the vacuum state. When the
elementary \ \ \ quarks are in the vacuum state, their colors = $\mathbb{B}$
= s = I = I$_{z}$ = Q = m = 0. Although they cannot be seen, as the physical
vacuum background, they really exist in the vacuum state and there is not
any other kind of quark in the vacuum state.

3). When an elementary quark gets enough energy, it may be excited into a
current quark. All quarks inside baryons and mesons are the excited states
of the elementary quark $\epsilon $. The u-quark and the c-quark are the
excited states of the elementary quark $\epsilon _{u}$; and the d-quark, the
s-quark and the b-quark are the excited states of the elementary quark $%
\epsilon _{d}$. The masses and flavors of the quarks can deduce from the
elementary quarks using (\ref{Qk-Mass}) and (\ref{flavor}).

4). Since all quarks are the excited states of the elementary quarks ($%
\epsilon _{\text{u}}$ and $\epsilon _{\text{d}}$) and the two elementary
quarks ($\epsilon _{\text{u}}$ and $\epsilon _{\text{d}}$) have SU(2)$_{%
\text{f}}$ symmetry, we can easily understand that SU(3)$_{\text{f}}$, SU(4)$%
_{\text{f}}$ and SU(5)$_{\text{f}}$ are natural extensions of SU(2)$_{\text{f%
}}$.\ The physical foundation of the extensions is that all quarks inside
hadrons are the excited states of the same elementary quark family $\epsilon 
$.

5). Using the sum laws and the binding energy formula of baryons, in terms
of SU(4) and qqq baryon model, we have deduced the masses and quantum
numbers of the important baryons from the deduced quarks in Table 2. At the
same time, using the sum laws and the binding energy formula of mesons, in
terms of q$\overline{\text{q}}$ meson model, we have also deduced the masses
and quantum numbers of the important mesons from the deduced quarks in Table
2. The deduced quantum numbers of the baryons and mesons match the
experimental results \cite{Baryon} and \cite{Meson} exactly, and the deduced
masses of the baryons and mesons are 98\% consistent with experimental
results. This case might show that the deduced masses of the quarks could be
indeed correct.

6). This paper not only deduce the masses and the\ quantum numbers
(including the flavors) of the quarks, the important baryons and mesons, but
also provides a physical foundation for the flavor symmetry (SU(3)$_{\text{f}%
},$SU(4)$_{\text{f}}$ and SU(5)$_{\text{f}}$) on a phenomenological level.
Since the elementary quark number decreases to two, it can decrease the
number of the adjustable parameters of the Standard Model \cite%
{Theory-Particle}, and might simplify the calculations. In fact this paper
improves upon the Quark Model, making it more powerful and more reasonable.

\begin{center}
\bigskip \textbf{Acknowledgments}
\end{center}

I sincerely thank Professor Robert L. Anderson for his valuable advice. I
acknowledge\textbf{\ }my indebtedness to Professor D. P. Landau for his help
also. I would like to express my heartfelt gratitude to Dr. Xin Yu. I
sincerely thank Professor Kang-Jie Shi for his important advice.\ \ \ \

\ \ \ \ \ \ \ \ \ \ \ \ \ \ \ \ \ \ \ \ \ \ \ 

\end{document}